\documentclass[twocolumn]{aastex631}
\usepackage{enumitem}
\usepackage{longtable}  
\usepackage{array}
\usepackage{rotating}  

\setlist[enumerate]{itemsep=0pt,parsep=0pt}
\begin{document}

\title{A search of periodic variable stars in the LMC by JWST photometry}

\author{Jiyu Wang}
\affiliation{CAS Key Laboratory of Optical Astronomy, National Astronomical Observatories, Chinese Academy of Sciences, Beijing 100101, China}
\affiliation{School of Astronomy and Space Science, University of the Chinese Academy of Sciences, Beijing, 100049, China}

\author{Xiaodian Chen}
\affiliation{CAS Key Laboratory of Optical Astronomy, National Astronomical Observatories, Chinese Academy of Sciences, Beijing 100101, China}
\affiliation{School of Astronomy and Space Science, University of the Chinese Academy of Sciences, Beijing, 100049, China}
\affiliation{Institute for Frontiers in Astronomy and Astrophysics, Beijing Normal University, Beijing 102206, China}

\author{Jianxing Zhang}
\affiliation{CAS Key Laboratory of Optical Astronomy, National Astronomical Observatories, Chinese Academy of Sciences, Beijing 100101, China}
\affiliation{School of Astronomy and Space Science, University of the Chinese Academy of Sciences, Beijing, 100049, China}

\author{Ziming Yan}
\affiliation{CAS Key Laboratory of Optical Astronomy, National Astronomical Observatories, Chinese Academy of Sciences, Beijing 100101, China}
\affiliation{School of Astronomy and Space Science, University of the Chinese Academy of Sciences, Beijing, 100049, China}

\author{Shu Wang}
\affiliation{CAS Key Laboratory of Optical Astronomy, National Astronomical Observatories, Chinese Academy of Sciences, Beijing 100101, China}

\author{Licai Deng}
\affiliation{CAS Key Laboratory of Optical Astronomy, National Astronomical Observatories, Chinese Academy of Sciences, Beijing 100101, China}
\affiliation{School of Astronomy and Space Science, University of the Chinese Academy of Sciences, Beijing, 100049, China}

\correspondingauthor{Xiaodian Chen}
\email{chenxiaodian@nao.cas.cn}

\begin{abstract}
Based on high-resolution near-infrared photometric data from the James Webb Space Telescope (JWST) targeting the Large Magellanic Cloud (LMC), this study attempts to evaluate the feasibility and sensitivity limits of variable star detection in crowded stellar fields. Through light curve analysis, we identified a total of 304 periodic variable stars, including 71 EW-type eclipsing binaries, 7 EA-type eclipsing binaries, 177 rotational variables, 38 $\delta$ Scuti (DSCT) stars, and 12 RR Lyrae stars. Period--luminosity relations (PLRs) were derived for EW-type eclipsing binaries, DSCT stars, and RR Lyrae stars. The PLRs for EW-type and RR Lyrae stars are in good agreement with previous studies, while the PLR zero point for DSCT stars appears systematically fainter by approximately 0.15--0.30 mag. Our PLRs exhibit low dispersion and are minimally affected by crowding. We analyzed the capability of JWST archival data to detect low-amplitude variables and found that only stars with amplitudes greater than approximately 0.05 mag can be reliably detected. Through simulations, we quantified how increasing the number of photometric epochs improves the detectability of low-amplitude, low signal-to-noise ratio variables. Despite current limitations in observational cadence, JWST demonstrates unique advantages in detecting short-period eclipsing binaries, rotational variables, and high-amplitude pulsators. Its exceptional spatial resolution enables high-precision PLR calibrations, offering new opportunities for future studies in variable star astrophysics and extragalactic distance measurements.

\end{abstract}


\section{Introduction}\label{sec:intro}
Variable stars, as key probes for stellar evolution dynamics, galaxy chemical evolution, and cosmic distance scale calibration, have long been limited in their detection and classification due to the spatial resolution, photometric sensitivity, and time-domain coverage of observational techniques \citep{2010ARA&A..48..673F}. Significant extinction effects ($A_{V} > 5$ mag) in the optical band, especially in dust-obscured regions such as galactic nuclei or star-forming regions, lead to systematic failures in detecting variable stars in these high-extinction areas \citep{2010NewA...15..433M, 2018ApJS..237...28C}. While traditional near-infrared photometry (e.g., the F160W band of HST/WFC3) partially mitigates the extinction effects, it still faces severe crowding noise in dense fields with angular separations smaller than 0.2\arcsec, restricting the detection efficiency of faint variable stars ($m_{\text{AB}} > 26$ mag). 

The James Webb Space Telescope (JWST) Near-Infrared Camera (NIRCam), with its diffraction-limited resolution ($< 0.1\arcsec$ at 2 $\mu$m) and broad wavelength coverage (1--5 $\mu$m), has, for the first time, enabled high-precision photometry in extremely crowded fields of the Large Magellanic Cloud (LMC), such as the core of star clusters \citep{2022MNRAS.517..484N}. Its sensitivity is sufficient to detect faint targets down to $\sim$ 29 mag, theoretically revealing variable stars that were previously inaccessible with conventional infrared facilities \citep{2022ApJ...936L..14P}.

However, the JWST was not primarily designed for time-domain astronomy. Its observation time is subject to strict scheduling constraints, making continuous monitoring (such as daily sampling for over 30 days) difficult. This results in gaps in the time-domain baseline and incomplete coverage of periods \citep{2023PASP..135d8001R}. The sparse and uneven sampling can severely contaminate the power spectrum of periodic signals with window function noise, especially for unknown or low-amplitude variables, where the period identification rate may fall to the level of random noise \citep{2018ApJS..236...16V}. Although JWST has demonstrated exceptional performance in exoplanet atmospheric spectroscopy and time-domain observations (e.g., transiting events, \citealt{2023PASP..135a8002E}), existing studies have primarily focused on static photometry (e.g., the deep analysis of the WLM galaxy by \citet{2023ApJS..268...15W}), and the quantification of time-domain systematics (such as detector gain drift and position-dependent biases in multi-epoch photometry) remains insufficient.

The LMC, as a nearby spiral galaxy (distance $\sim$ 50 kpc), offers unique advantages for variable star research: its low interstellar extinction ($A_{V} \sim 0.3$ mag) allows for clear imaging of dust-obscured regions, and the diverse stellar populations (ranging from young OB stars to older red giant cluster stars) along with high-precision color-magnitude diagrams (CMD) provide independent multi-dimensional constraints on the physical parameters (mass, metallicity, evolutionary stage) of variable stars, independent of light curve (LC) analysis \citep{2009AJ....138.1243H}. However, traditional near-infrared surveys (such as VISTA/VMC) are limited by resolution ($\sim 0.8\arcsec$) and sensitivity ($K_{s} < 20$ mag), detecting only the brighter variable stars in the red giant stars of the LMC, with coverage for faint targets (such as main sequence variables) under 1\% \citep{2011A&A...527A.116C,2014A&A...562A..32C}. The deep-field capabilities of JWST fill this observational gap, but its sparse sampling nature presents a fundamental challenge for detecting unknown periodic variables.

Our study aims to explore the feasibility of detecting unknown variable stars in crowded fields of the LMC using JWST/NIRCam. By utilizing high-precision PSF modeling with DOLPHOT to eliminate contamination from neighboring stars \citep{2000PASP..112.1383D,2016ascl.soft08013D} and correlating multi-band LCs, we aim to accurately identify variable stars. Additionally, we seek to quantify the ability of JWST/NIRCam in detecting low signal-to-noise ratio (SNR) and unknown periodic variables. This work not only aims to uncover the hidden diversity of variable stars in the LMC, but also provides critical insights for optimizing future JWST time-domain observational strategies (such as prioritizing continuous observation windows).

The structure of the paper is as follows. Section \ref{sec:data} delineates the data acquisition methodology and the rigorous criteria applied to curate a clean stellar sample. Section \ref{sec:result} details the systematic identification of robust variable star candidates through LC diagnostics. In Section \ref{sec:plrs}, we fit the period-luminosity relations (PLR) of all confirmed variable stars in the sample. Section \ref{sec:discuss} employs Monte Carlo simulations to quantify the detection thresholds of variability amplitudes achievable under the current data quality, explicitly parameterized by SNR. Finally, we provide a summary in Section \ref{conclude}.

\begin{figure*}[ht]
    \centering
    \includegraphics[width=1\linewidth]{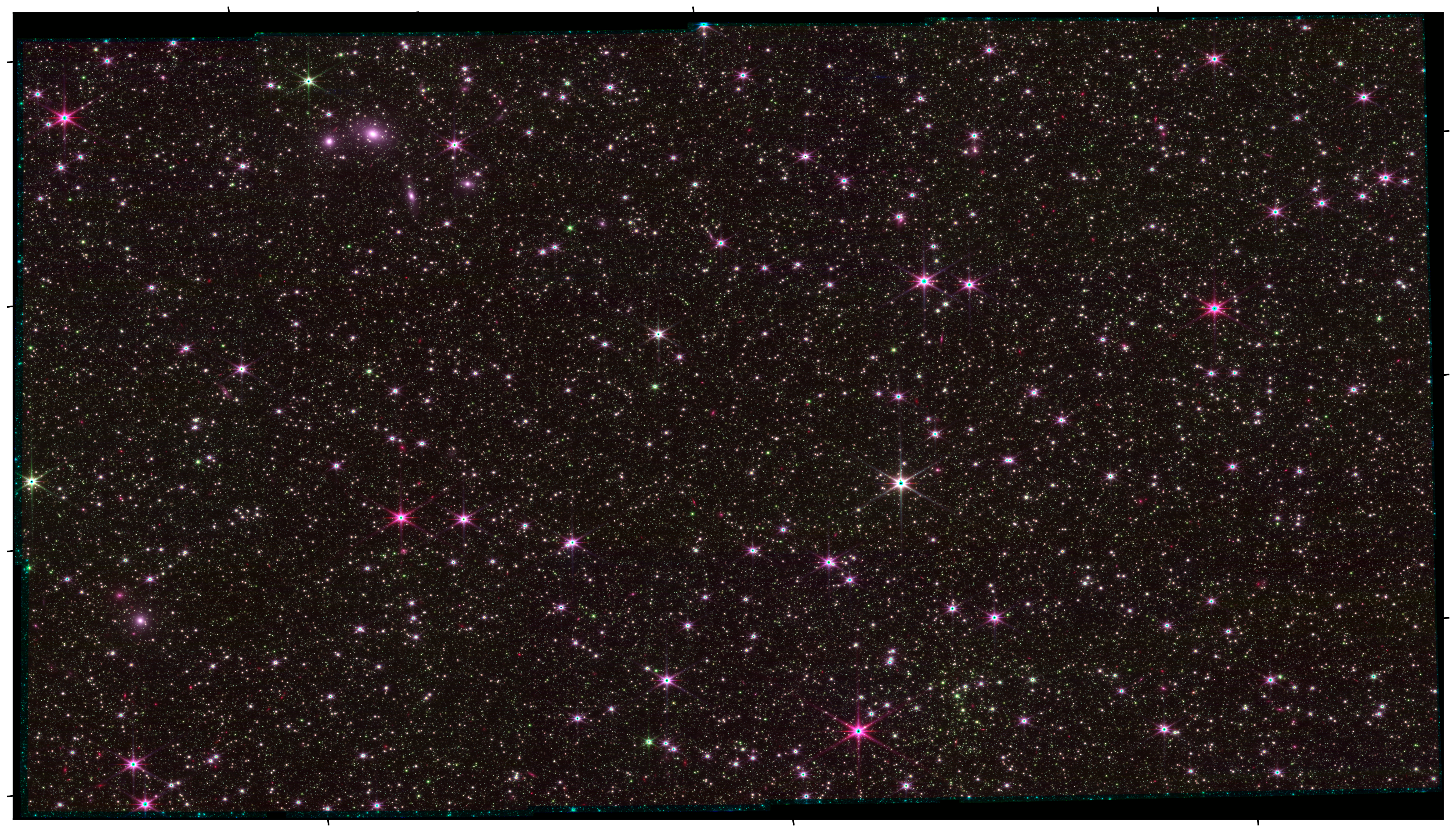}
    \caption{The figure shows a JWST/NIRCam composite image in six filters (F070W, F150W, F200W, F277W, F356W, F444W) of a $\sim6.2\arcmin \times 3.6\arcmin$ field in the LMC, representing the primary region analyzed in this study.
}
    \label{fig:1}
\end{figure*}

\begin{figure}[ht]
    \centering
    \includegraphics[width=1\linewidth]{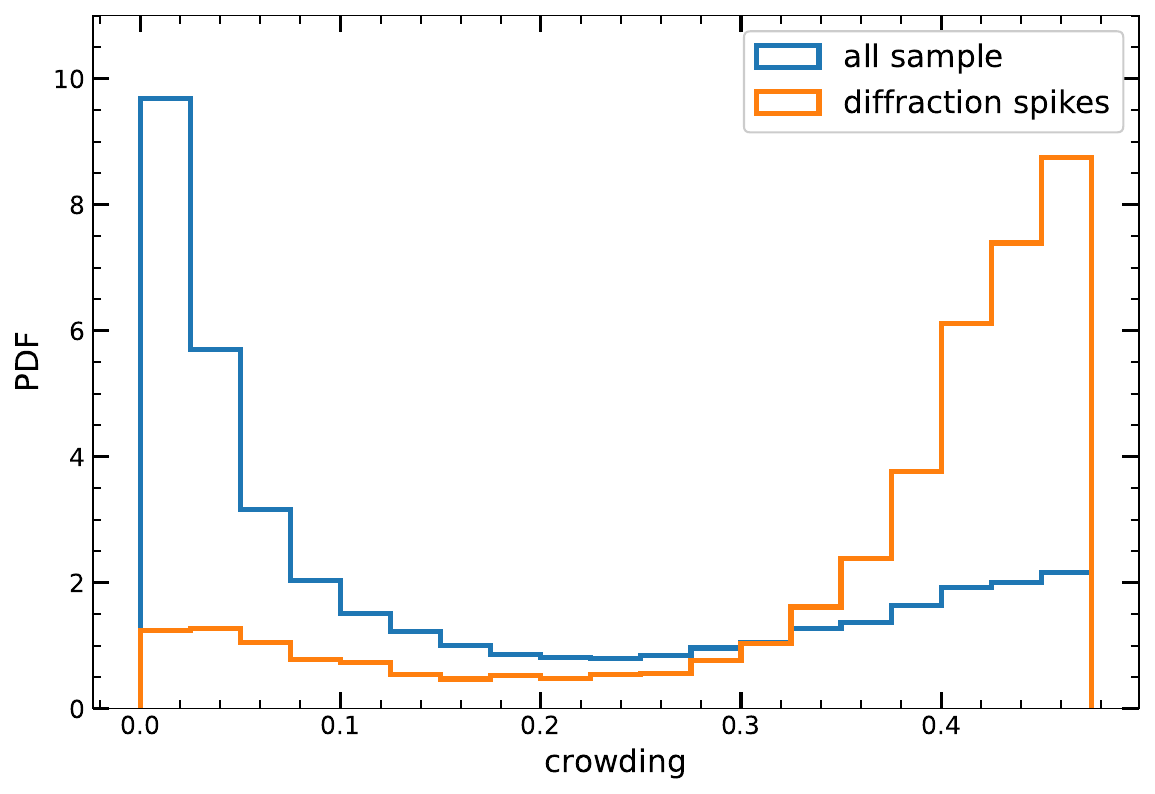}
    \caption{Distribution of crowding parameters for spurious sources from diffraction spikes and the overall sample. The crowding distributions exhibit a pronounced disparity between spurious sources and normal stars, with spurious sources predominantly clustered above 0.3 while typical stars are primarily concentrated below 0.2.}
    \label{fig:2}
\end{figure}

\section{Data}\label{sec:data}
The survey region encompasses the combined coverage of multiple JWST observational programs (Program IDs: 1069, 1072, 1073, 1074, 1473, 1476, 1477, 4447, and 6627) whose central coordinates lie within a 3 arcmin radius centered at $\alpha = 80.49031^\circ$, $\delta = -69.498156^\circ$ (J2000). Figure~\ref{fig:1} illustrates the primary region analyzed in this study, highlighting the densely populated stellar field in the LMC.

\begin{enumerate}
    \item Filters selection: F070W, F090W, F115W, F150W, F200W, F277W, F356W and F444W.
    \item Exposure threshold: Photometric data was required to have cumulative exposure counts $\geq$ 40 per filter. 
    The F090W and F115W filters were excluded due to insufficient exposures.
\end{enumerate}

Some of the data used in this study were obtained from the Mikulski Archive for Space Telescopes (MAST): \dataset[doi:10.17909/p2pc-g715]{https://doi.org/10.17909/p2pc-g715}.

\subsection{photometry}\label{sec:photo}
In the photometric processing of this study, we performed photometric analysis on the screened observational data using the \textsc{dolphot} package. The procedure requires a reference image and multiple science images as inputs, along with configurations for over forty critical parameters including the point spread function (PSF) model coefficients, local background fitting radius, and SNR thresholds. The PSF-related parameter settings strictly follow the guidelines described in \citet{2024ApJS..271...47W}.
Given the multi-epoch photometric data originated from distinct observing proposals, no stage3 co-added image covering the entire photometric field was available. Consequently, we implemented a matched single-exposure photometric strategy, where science images utilize stage2-processed single-exposure calibrated files (\texttt{\_cal.fits}), while reference images adopt stage2 co-added images (\texttt{\_i2d.fits}) corresponding to the same exposure. This approach can maintain the spatial variation traits of the PSF unique to each exposure, prevent systematic smoothing artifacts that might occur during stage3 image combination procedures, and improve computational efficiency by minimizing image resampling operations, all while ensuring consistency in measurements across individual exposures.

\subsection{data selection}\label{sec:filter}
After finishing the photometric processing, the results for each scientific target were saved as separate ASCII files. Since the DOLPHOT photometric software generates comprehensive catalogs of all detected sources within the imaging data, we established a rigorous set of quality standards to extract high-purity and high-quality photometric data. Our final selection criteria are described as follows:

\begin{enumerate}
    \item {SNR} $\geq$ 5
    \item Sharpness$^2$ $\leq$ 0.01
    \item Crowding $\leq$ 0.3
    \item Photometry quality flag $<$ 1
    \item Object type $\leq$ 2
    \item Roundness $\leq$ 2
\end{enumerate}

\begin{figure*}[ht]
    \centering
    \includegraphics[width=1\linewidth]{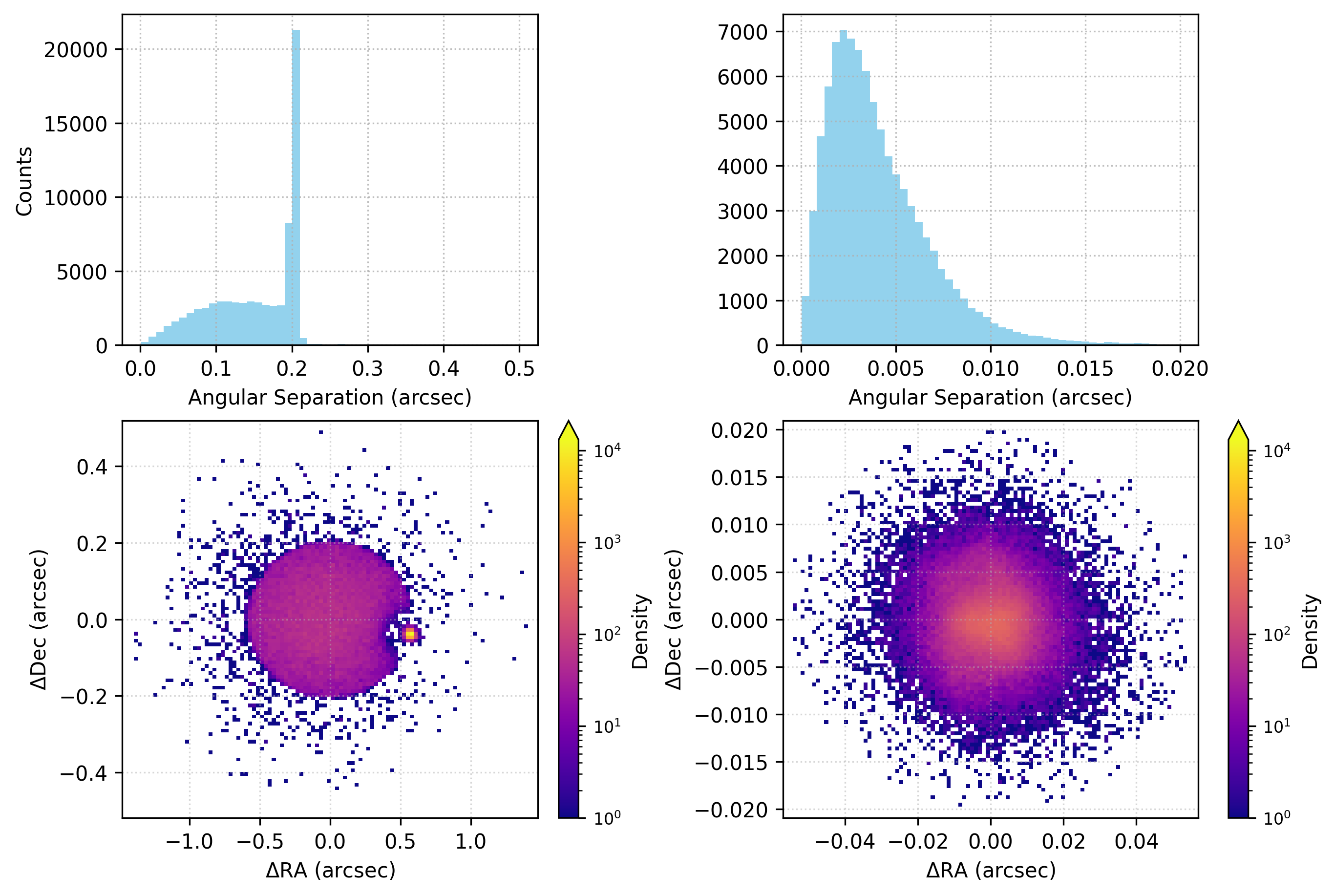}
    \caption{The figure presents an example of the cross-matching results between the reference catalog and the catalog to be corrected, including a histogram of angular separations and scatter plots of right ascension and declination residuals. The left panels show the distribution before astrometric calibration, where correctly matched sources form a high-density cluster in the residual plot. The right panels display the post-correction distribution, with most matched sources exhibiting angular separations less than 0.01\arcsec.}
    \label{fig:3}
\end{figure*}

\begin{figure*}[ht]
    \centering
    \includegraphics[width=1\linewidth]{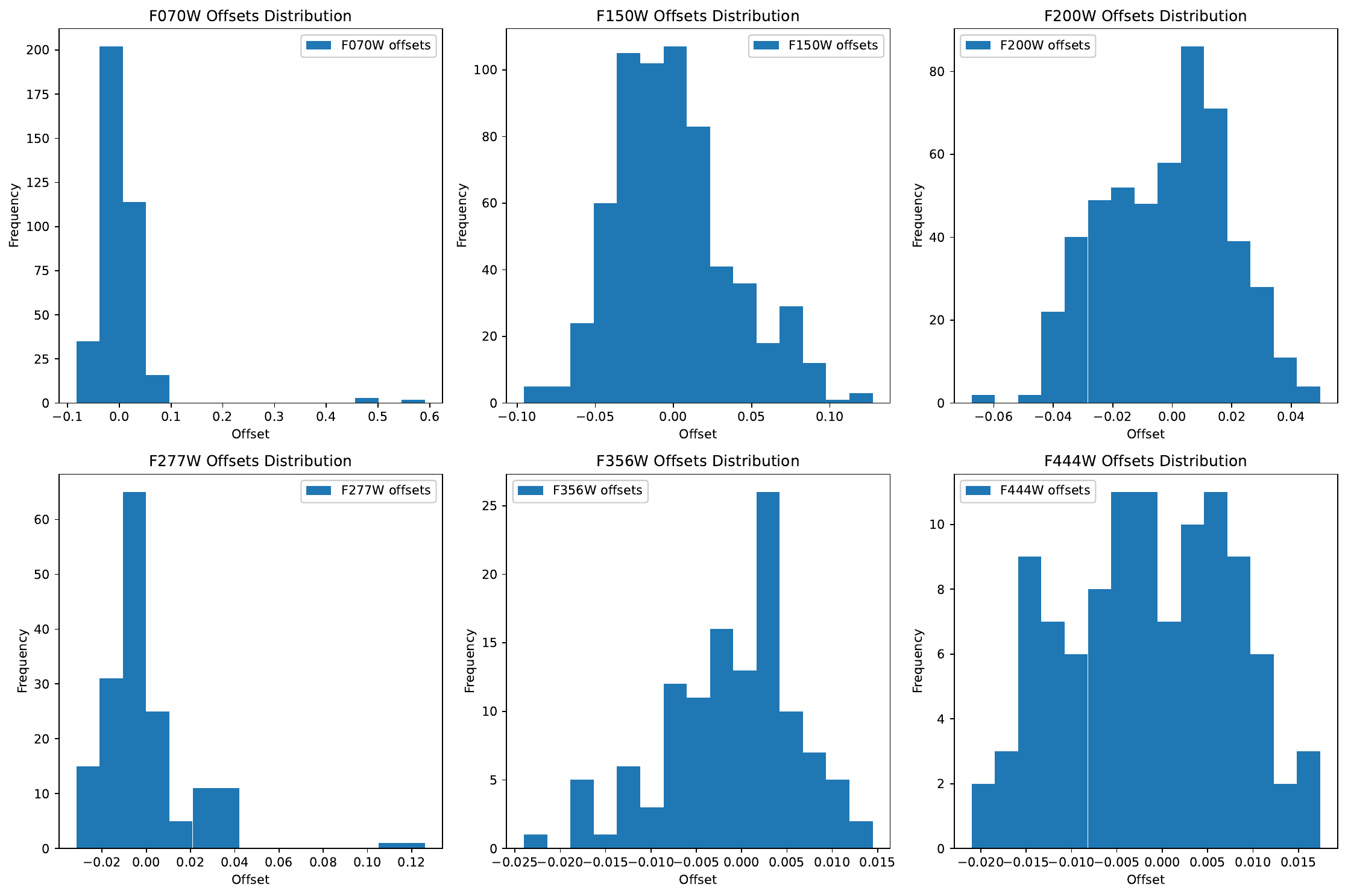}
    \caption{The figure illustrates the distribution of zero-point corrections across different filters, with short-wavelength bands exhibiting more pronounced corrections compared to long-wavelength bands. Notably, anomalous offsets deviating from the overall trend are observed in the F070W and F277W bands. Given that each detector module contains a calibration star sample exceeding 200 sources – satisfying statistical robustness criteria – these anomalies are unlikely to originate from photometric errors of individual sources. }
    \label{fig:4}
\end{figure*}

Compared to the selection criteria adopted by \citet{2024ApJS..271...47W}, this study imposes stricter thresholds for crowding and photometry quality flags while additionally introducing roundness as a filtering parameter. This optimization benefits from the high-quality photometric data of LMC and aligns with the precision requirements for variable star detection.

About the SNR, this study adopts a threshold of SNR $\geq 5$, consistent with the standard set by \citet{2024ApJS..271...47W} but lower than the criteria applied by \citet{2023RNAAS...7...23W} in their JWST photometric data processing. This choice is primarily motivated by considerations of variable star detection completeness. Although low-SNR stars exhibit larger photometric errors—potentially obscuring periodic signals from low-amplitude variables (particularly pulsating stars, whose amplitudes exhibit wavelength-dependent attenuation)—it remains possible to detect high-amplitude rotating variables and eclipsing binaries.

Regarding the setting of the crowding parameter, this study strictly defines the threshold as \textit{crowding} $\leq 0.3$, which is more restrictive compared to the criterion of \textit{crowding} $\leq 0.5$ adopted by \cite{2024ApJS..271...47W}. Directly adopting the criterion of \textit{crowding} $\leq 0.5$ from \cite{2024ApJS..271...47W} would be insufficient to exclude spurious sources such as diffraction spikes in the sample. As shown in Figure.~\ref{fig:2}, the crowding parameter of diffraction spikes is concentrated in the range of $0.3$--$0.5$, while the crowding values for normal stars in the overall sample primarily fall below $0.2$. Therefore, adopting a crowding threshold of $\leq 0.3$ aids in obtaining a sample of higher purity.

\subsection{coordinate correction}\label{sec:correct}

We introduce a coordinate correction approach leveraging cross-identification to mitigate systematic offsets in the World Coordinate System (WCS) included in JWST data products. Systematic errors in the Guide Star Catalog mainly stem from guide star misidentification and uncertainties in spacecraft roll angle measurements \citep{jwstDoc3}. The statistical analysis reveals a radial deviation of 0.3\arcsec at a 1$\sigma$ confidence level, with some instances surpassing 1\arcsec \citep{jwstDoc6}.

The calibration process uses the star catalog with the highest source density as the reference frame, and the data from other observation epochs are spatially aligned through coordinate transformations. During the cross-validation process, two typical cases were identified: (1) Matched pairs corresponding to genuine astrophysical sources exhibit significant clustering in the two-dimensional residual parameter space $(\Delta \delta, \Delta \alpha)$; (2) Randomly matched pairs lacking physical associations display uniform residual distributions.  As illustrated in the left panel of Figure~\ref{fig:3}, the angular separation distribution exhibits a prominent peak near 0.2\arcsec, corresponding to localized density maxima in the $(\Delta \delta, \Delta \alpha)$ phase space (color-mapped visualization). Leveraging this characteristic, we utilize the Density-Based Spatial Clustering of Applications with Noise (DBSCAN) algorithm to extract high-density regions in the residual space and derive coordinate transformation parameters via least-squares fitting. Post-calibration residual distributions, shown in the right panel of Figure~\ref{fig:3}, demonstrate that angular separations for the majority of matched sources are reduced to below 0.01\arcsec.

\subsection{zero-point correction}\label{sec:standard}
The JWST NIRCAM exhibits minor sensitivity variations between observational epochs, while source positional offsets caused by insufficient image registration accuracy across different visits may introduce systematic photometric errors. Additionally, zero-point offsets exist between detector modules, typically below 0.05 mag \citep{2024ApJS..271...47W}, but potentially exceeding 0.2 mag under specific conditions \citep{2022RNAAS...6..191B}. Consequently, zero-point correction plays a dual critical role in variable source detection: it mitigates systematic errors to reduce noise in LCs while enhancing SNR and data reliability, and ensures photometric consistency across visits \citep{2019A&A...630A..92B}.

In this study, standard stars were selected through two criteria: (1) availability of $\geq$ 10 independent photometric measurements, and (2) photometric scatter satisfying $\sigma < 2\,\langle\Delta m\rangle$, where $\Delta m$ denotes the mean photometric uncertainty. This eliminates outliers and potential variable sources through multi-epoch averaging. For each detector module, we computed the median deviation between single-epoch photometry and the long-term average magnitude of standard stars, adopting this median as the module-specific zero-point correction.

Figure~\ref{fig:4} illustrates the statistical distribution of zero-point corrections across all bands. While corrections for long-wavelength bands $(\lambda > 2.36\ \mu m)$ generally remain below 0.02 mag, significantly larger adjustments are observed in short-wavelength bands, underscoring the critical necessity of zero-point correction for shortwave NIRCam observations.

\begin{deluxetable*}{lcccccccccccc}
\tablecaption{LMC periodic variable star catalog based on JWST\label{tab:1}}
\tablehead{
\colhead{ID} & \colhead{RA (J2000)} & \colhead{Dec (J2000)} & \colhead{F150W} &
\colhead{$F150W_{\rm err}$} & \colhead{P (days)} & \colhead{FAP} &
\colhead{Amp (mag)} & \colhead{RMSE} & \colhead{$R^2$} & \colhead{Type} &
\colhead{$R_{21}$} & \colhead{$\phi_{21}$($\pi$)}
}
\startdata
1   & 80.47703 & -69.503902 & 20.219 & 0.003          & 0.0741668    & 4.5E-9  & 0.182            & 0.022 & 0.907      & DSCT & 0.627 & 2.502 \\
38  & 80.4304  & -69.510344 & 23.192 & 0.011          & 0.1030841    & 2.6E-12 & 0.607            & 0.08  & 0.854      & EW   & 0.29  & 1.975 \\
39  & 80.48846 & -69.488704 & 22.816 & 0.008          & 0.105515     & 4.7E-17 & 0.548            & 0.055 & 0.907      & EW   & 0.324 & 2.111 \\
40  & 80.48315 & -69.498244 & 22.85  & 0.006          & 0.1076266    & 2.7E-19 & 0.395            & 0.044 & 0.915      & EW   & 0.178 & 2.099 \\
41  & 80.48076 & -69.500459 & 22.229 & 0.005          & 0.1094874    & 2.9E-13 & 0.28             & 0.037 & 0.857      & EW   & 0.211 & 1.99  \\
42  & 80.52676 & -69.516807 & 22.454 & 0.006          & 0.1175075    & 2.1E-10 & 0.287            & 0.035 & 0.883      & EW   & 0.19  & 2.121 \\
43  & 80.54841 & -69.497576 & 22.415 & 0.005          & 0.1206538    & 2.1E-14 & 0.517            & 0.031 & 0.97       & EW   & 0.247 & 1.961 \\
44  & 80.4748  & -69.495366 & 21.826 & 0.007          & 0.1261665    & 1.8E-13 & 0.385            & 0.047 & 0.886      & EW   & 0.339 & 1.91  \\
45  & 80.50329 & -69.497994 & 22.18  & 0.006          & 0.1267689    & 1.2E-16 & 0.398            & 0.047 & 0.897      & EW   & 0.273 & 1.941 \\
46  & 80.51625 & -69.474229 & 21.927 & 0.006          & 0.1272398    & 6.2E-9  & 0.263            & 0.039 & 0.808      & EW   & 0.1   & 1.857 \\
47  & 80.53764 & -69.482711 & 22.309 & 0.005          & 0.1282954    & 2.1E-15 & 0.688            & 0.036 & 0.981      & EW   & 0.416 & 1.993 \\
48  & 80.53661 & -69.477373 & 22.192 & 0.006          & 0.1288176    & 4.4E-18 & 0.414            & 0.043 & 0.932      & EW   & 0.196 & 1.79  \\
49  & 80.51416 & -69.484409 & 22.219 & 0.004          & 0.1296082    & 1.6E-15 & 0.389            & 0.028 & 0.941      & EW   & 0.272 & 2.046 \\
50  & 80.51837 & -69.519215 & 22.14  & 0.007          & 0.1317063    & 2.0E-9  & 0.34             & 0.04  & 0.894      & EW   & 0.368 & 2.096 \\
51  & 80.51538 & -69.491573 & 21.933 & 0.007          & 0.1330268    & 7.1E-17 & 0.412            & 0.048 & 0.898      & EW   & 0.252 & 2.167 \\
52  & 80.48502 & -69.529347 & 21.879 & 0.006          & 0.1386334    & 4.1E-12 & 0.423            & 0.034 & 0.949      & EW   & 0.209 & 1.617 \\
53  & 80.47881 & -69.510834 & 21.558 & 0.007          & 0.1409486    & 5.5E-12 & 0.344            & 0.048 & 0.849      & EW   & 0.255 & 2.153 \\
54  & 80.51536 & -69.517658 & 21.756 & 0.005          & 0.1432884    & 2.7E-19 & 0.397            & 0.036 & 0.946      & EW   & 0.26  & 1.969 \\
55  & 80.4646  & -69.498904 & 21.46  & 0.006          & 0.1433882    & 4.2E-9  & 0.21             & 0.044 & 0.695      & EW   & 0.182 & 2.031 \\
56  & 80.49107 & -69.498531 & 21.499 & 0.003          & 0.1451683    & 7.6E-23 & 0.276            & 0.023 & 0.94       & EW   & 0.221 & 1.935 \\
57  & 80.49661 & -69.49895  & 21.414 & 0.003          & 0.1484179    & 3.1E-13 & 0.126            & 0.025 & 0.751      & EW   & 0.049 & 2.949 \\
58  & 80.49268 & -69.496674 & 21.823 & 0.005          & 0.1492539    & 3.0E-15 & 0.547            & 0.039 & 0.94       & EW   & 0.32  & 2.018 \\
59  & 80.49759 & -69.499632 & 21.576 & 0.003          & 0.1500697    & 1.6E-24 & 0.3              & 0.026 & 0.949      & EW   & 0.219 & 2.113 \\
60  & 80.45563 & -69.474669 & 21.612 & 0.006          & 0.1501264    & 7.1E-12 & 0.417            & 0.033 & 0.953      & EW   & 0.139 & 2.026 \\
61  & 80.51585 & -69.482207 & 21.627 & 0.006          & 0.1501991    & 1.8E-18 & 0.38             & 0.039 & 0.93       & EW   & 0.15  & 1.971 \\
62  & 80.48945 & -69.491393 & 21.723 & 0.004          & 0.1515267    & 6.1E-19 & 0.494            & 0.032 & 0.966      & EW   & 0.368 & 1.957 \\
64  & 80.54501 & -69.514314 & 21.43  & 0.005          & 0.1569988    & 2.6E-13 & 0.363            & 0.025 & 0.963      & EW   & 0.143 & 2.203 \\
65  & 80.50197 & -69.498821 & 21.14  & 0.004          & 0.1621966    & 1.2E-13 & 0.159            & 0.033 & 0.771      & EW   & 0.194 & 1.004 \\
66  & 80.49907 & -69.47365  & 21.41  & 0.006          & 0.1623324    & 3.5E-6  & 0.219            & 0.036 & 0.728      & EW   & 0.217 & 1.871 \\
67  & 80.48561 & -69.528405 & 21.123 & 0.004          & 0.1675978    & 1.6E-14 & 0.296            & 0.024 & 0.956      & EW   & 0.163 & 2.109 \\
68  & 80.50553 & -69.501186 & 21.806 & 0.005          & 0.1692361    & 1.9E-6  & 0.331            & 0.034 & 0.806      & EW   & 0.338 & 1.837 \\
69  & 80.48343 & -69.52792  & 21.045 & 0.005          & 0.1746695    & 2.8E-16 & 0.334            & 0.032 & 0.931      & EW   & 0.102 & 1.926 \\
70  & 80.45733 & -69.519904 & 21.194 & 0.006          & 0.1756267    & 9.9E-11 & 0.283            & 0.036 & 0.866      & EW   & 0.014 & 1.21  \\
71  & 80.42064 & -69.516618 & 21.427 & 0.008          & 0.1771504    & 9.8E-14 & 0.607            & 0.047 & 0.955      & EW   & 0.264 & 1.96  \\
72  & 80.5313  & -69.51231  & 21.067 & 0.004          & 0.1784391    & 1.1E-13 & 0.276            & 0.023 & 0.939      & EW   & 0.225 & 1.933 \\
73  & 80.45906 & -69.488627 & 21.042 & 0.005          & 0.1799475    & 2.6E-15 & 0.276            & 0.029 & 0.944      & EW   & 0.191 & 2.444 \\
74  & 80.53507 & -69.502213 & 21.114 & 0.004          & 0.181607     & 1.9E-18 & 0.348            & 0.026 & 0.963      & EW   & 0.146 & 2.305 \\
75  & 80.44995 & -69.474427 & 21.028 & 0.007          & 0.1827129    & 1.6E-10 & 0.373            & 0.032 & 0.958      & EW   & 0.127 & 2.212 \\
76  & 80.47774 & -69.502472 & 21.02  & 0.004          & 0.1865572    & 1.4E-22 & 0.437            & 0.029 & 0.955      & EW   & 0.136 & 2.011 \\
77  & 80.48059 & -69.480204 & 21.121 & 0.007          & 0.1878518    & 1.0E-10 & 0.417            & 0.045 & 0.887      & EW   & 0.201 & 2.155 \\
78  & 80.55126 & -69.48747  & 21.283 & 0.004          & 0.2253226    & 2.2E-12 & 0.422            & 0.025 & 0.874      & EW   & 0.359 & 2.023 \\
79  & 80.5499  & -69.492934 & 20.652 & 0.005          & 0.2285189    & 4.6E-15 & 0.345            & 0.032 & 0.926      & EW   & 0.115 & 1.965 \\
80  & 80.53491 & -69.493899 & 20.44  & 0.003          & 0.2333549    & 5.1E-24 & 0.297            & 0.02  & 0.963      & EW   & 0.192 & 2.019 \\
81  & 80.52217 & -69.51915  & 20.475 & 0.005          & 0.2657121    & 5.9E-10 & 0.179            & 0.03  & 0.817      & EW   & 0.124 & 1.055 \\
82  & 80.46355 & -69.482672 & 19.373 & 0.003          & 0.3284697    & 2.0E-10 & 0.159            & 0.022 & 0.822      & EW   & 0.142 & 2.011 \\
83  & 80.41861 & -69.521674 & 19.837 & 0.005          & 0.3465736    & 8.1E-15 & 0.344            & 0.03  & 0.943      & EW   & 0.142 & 1.902 \\
84  & 80.46269 & -69.505615 & 22.655 & 0.006          & 0.1131617    & 4.5E-10 & 0.446            & 0.038 & 0.941      & EW   & 0.429 & 2.179 \\
85  & 80.52278 & -69.519601 & 19.639 & 0.004          & 0.3421849    & 4.3E-16 & 0.319            & 0.024 & 0.954      & EW   & 0.17  & 2.112 \\
293 & 80.46195 & -69.512171 & 18.46  & 0.003          & 0.3132741    & 4.4E-11 & 0.16             & 0.022 & 0.82       & RR   & 0.157 & 2.39  \\
294 & 80.56921 & -69.495232 & 18.323 & 0.004          & 0.3596203    & 1.4E-8  & 0.135            & 0.021 & 0.833      & RR   & 0.238 & 1.816 \\
295 & 80.47905 & -69.518617 & 18.24  & 0.003          & 0.5608823    & 2.8E-12 & 0.202            & 0.02  & 0.895      & RR   & 0.14  & 2.418
\enddata
\tablecomments{ Only a subset of rows and columns is shown here for illustration; the full table is available in machine-readable format.}
\end{deluxetable*}

\begin{table*}[ht]
\centering
\caption{Multi-eopch photometry for the variable stars listed in Table~\ref{tab:1}.}
\label{tab:2}
\begin{tabular}{ccccccc}
\hline
ID & RA & Dec & Filter & Mag & ${\rm Mag_{err}}$ & Obs. Time \\
   & (deg) & (deg) &  & (mag) & (mag) & (MJD) \\
\hline
129 &  80.44302 &-69.494037 &F070W & 24.682 & 0.083& 60159.2283645\\
129 &  80.44302 &-69.494037 &F070W & 24.707 & 0.086& 59696.4347108\\
129 &  80.44302 &-69.494037 &F070W & 24.806 & 0.087& 60193.4235530\\
129 &  80.44302 &-69.494037 &F070W & 24.591 & 0.082& 60159.1508690\\
129 &  80.44302 &-69.494037 &F070W & 24.548 & 0.049& 60625.1514348\\
129 &  80.44302 &-69.494037 &F070W & 24.790 & 0.089& 59696.3243272\\
129 &  80.44302 &-69.494037 &F070W & 24.605 & 0.079& 59696.4325982\\
129 &  80.44302 &-69.494037 &F150W & 22.865 & 0.133& 59990.0595977\\
129 &  80.44302 &-69.494037 &F150W & 22.956 & 0.007& 59772.1716754\\
129 &  80.44302 &-69.494037 &F150W & 22.973 & 0.035& 59738.7914412\\
129 &  80.44302 &-69.494037 &F150W & 23.054 & 0.035& 60193.4093864\\
129 &  80.44302 &-69.494037 &F150W & 23.003 & 0.036& 60159.1367024\\
129 &  80.44302 &-69.494037 &F150W & 23.003 & 0.034& 59885.1883000\\
129 &  80.44302 &-69.494037 &F150W & 23.050 & 0.037& 60159.2140734\\
129 &  80.44302 &-69.494037 &F150W & 22.987 & 0.007& 59772.1935465\\
129 &  80.44302 &-69.494037 &F150W & 23.084 & 0.037& 59696.3314101\\
129 &  80.44302 &-69.494037 &F150W & 23.035 & 0.035& 60193.4116234\\
129 &  80.44302 &-69.494037 &F150W & 22.956 & 0.030& 59692.2779921\\
129 &  80.44302 &-69.494037 &F150W & 22.934 & 0.034& 60159.2163105\\
129 &  80.44302 &-69.494037 &F150W & 22.982 & 0.035& 59738.7935538\\
129 &  80.44302 &-69.494037 &F150W & 22.940 & 0.029& 59692.2728973\\
129 &  80.44302 &-69.494037 &F150W & 22.950 & 0.030& 59692.3028143\\
129 &  80.44302 &-69.494037 &F150W & 22.919 & 0.029& 59692.3054240\\
129 &  80.44302 &-69.494037 &F150W & 22.892 & 0.029& 59692.3080336\\
129 &  80.44302 &-69.494037 &F150W & 23.015 & 0.037& 59738.7956664\\
129 &  80.44302 &-69.494037 &F150W & 23.070 & 0.037& 59696.3292975\\
129 &  80.44302 &-69.494037 &F150W & 23.042 & 0.035& 59720.3573977\\
129 &  80.44302 &-69.494037 &F150W & 23.068 & 0.037& 59696.4419189\\
129 &  80.44302 &-69.494037 &F150W & 23.028 & 0.045& 59671.3040631\\
129 &  80.44302 &-69.494037 &F150W & 22.938 & 0.030& 59692.2753825\\
... &  ...&...&...&...&...&...\\
\hline
\end{tabular}

\vspace{2mm}
\parbox{0.9\linewidth}{\footnotesize \textbf{Note.} This table shows only a portion of the full dataset. The complete table is available in machine-readable format.}
\end{table*}

\begin{table}[htbp]
\centering
\caption{Parameters of PLRs} 
\label{tab:3}
\begin{tabular}{ccccc}
\toprule
\textbf{$a_\lambda$} & \textbf{$b_\lambda$} & \textbf{$\sigma$} & \text{mode} & \text{Filter} \\ 
\multicolumn{5}{c}{$m_\lambda = a_\lambda \times \log P + b_\lambda$} \\ 
\hline
\multicolumn{5}{c}{RR Lyrae} \\  
\hline
$-3.135 \pm 0.899$ & $-1.044 \pm 0.426$ & 0.15 & 1O    & F070W  \\
$-2.104 \pm 0.352$ & $-0.075 \pm 0.093$  & 0.13 & F    & F070W  \\
$-3.005 \pm 0.333$ & $-1.478 \pm 0.160$ & 0.06 & 1O    & F150W  \\
$-2.795 \pm 0.208$ & $-0.951 \pm 0.056$ & 0.08 & F    & F150W  \\
$-2.919 \pm 0.359$ & $-1.552 \pm 0.172$ & 0.06 & 1O    & F200W  \\
$-2.909 \pm 0.217$ & $-1.099 \pm 0.059$ & 0.08 & F    & F200W  \\
$-3.172 \pm 0.239$ & $-1.724 \pm 0.114$ & 0.04 & 1O    & F277W  \\
$-2.893 \pm 0.232$ & $-1.156 \pm 0.063$ & 0.09 & F   & F277W  \\
$-3.167 \pm 0.199$ & $-1.750 \pm 0.095$ & 0.04 & 1O    & F356W  \\
$-2.862 \pm 0.243$ & $-1.170 \pm 0.066$ & 0.09 & F    & F356W  \\
$-2.758 \pm 0.365$ & $-1.561 \pm 0.175$ & 0.07 & 1O    & F444W  \\
$-2.958 \pm 0.241$ & $-1.218 \pm 0.065$ & 0.09 & F    & F444W  \\
\hline
\multicolumn{5}{c}{DSCT} \\  
\hline
$-4.090 \pm 0.882$ & $-2.930 \pm 0.962$ & 0.28 & 1O    & F070W  \\
$-4.850 \pm 0.397$ & $-3.268 \pm 0.436$ & 0.26 & F    & F070W  \\
$-4.279 \pm 0.377$ & $-3.556 \pm 0.412$ & 0.12 & 1O    & F150W  \\
$-4.096 \pm 0.188$ & $-2.855 \pm 0.206$ & 0.12 & F   & F150W  \\
$-4.276 \pm 0.345$ & $-3.631 \pm 0.376$ & 0.11 & 1O    & F200W  \\
$-4.038 \pm 0.189$ & $-2.878 \pm 0.207$ & 0.13 & F   & F200W  \\
$-4.234 \pm 0.357$ & $-3.645 \pm 0.390$ & 0.12 & 1O    & F277W  \\
$-3.975 \pm 0.188$ & $-2.869 \pm 0.207$ & 0.12 & F   & F277W  \\
$-4.273 \pm 0.330$ & $-3.699 \pm 0.360$ & 0.11 & 1O    & F356W  \\
$-3.941 \pm 0.192$ & $-2.847 \pm 0.211$ & 0.12 & F   & F356W  \\
$-4.258 \pm 0.332$ & $-3.701 \pm 0.363$ & 0.11 & 1O    & F444W  \\
$-3.993 \pm 0.202$ & $-2.911 \pm 0.222$ & 0.13 & F   & F444W  \\
\hline
\multicolumn{5}{c}{EW} \\  
\hline
$-9.307 \pm 0.415$ & $-0.839 \pm 0.223$ & 0.32 & {}   & F070W \\
$-6.688 \pm 0.305$ & $-0.342 \pm 0.164$ & 0.23 & {}   & F150W \\
$-6.259 \pm 0.286$ & $-0.270 \pm 0.153$ & 0.22 & {}   & F200W \\
$-6.200 \pm 0.304$ & $-0.354 \pm 0.162$ & 0.22 & {}   & F277W \\
$-6.053 \pm 0.330$ & $-0.298 \pm 0.175$ & 0.24 & {}   & F356W \\
$-5.914 \pm 0.324$ & $-0.249 \pm 0.171$ & 0.23 & {}   & F444W \\
\hline
\end{tabular}
\end{table}

\begin{figure*}[ht]
    \centering
    \includegraphics[width=1\linewidth]{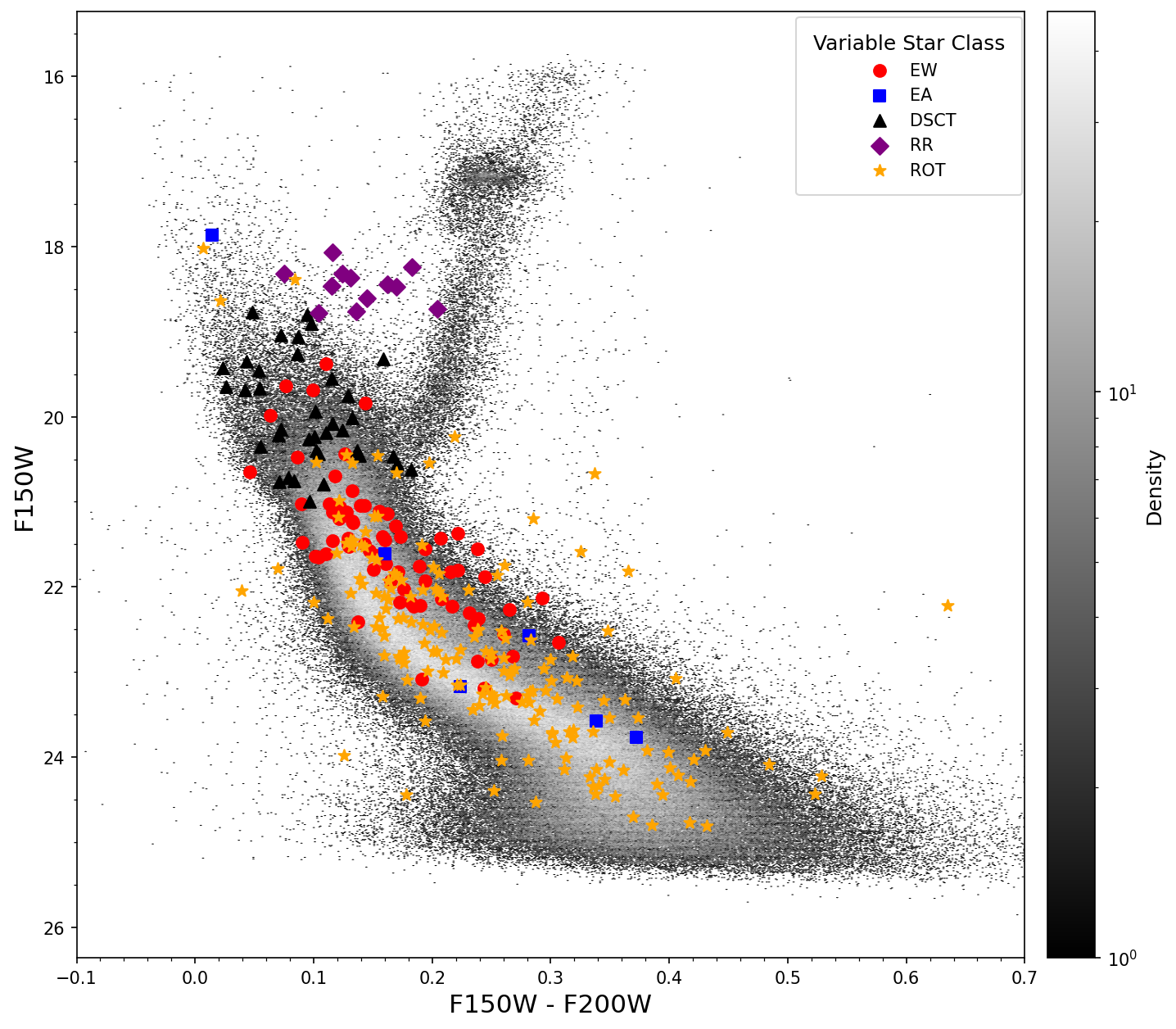}
    \caption{CMD of sources in the sample with more than five photometric epochs in both F150W and F200W bands. The density of sources is color-mapped, and confirmed variable stars are indicated by distinct symbols.}
    \label{fig:5}
\end{figure*}

\begin{figure*}[ht]
    \centering
    \includegraphics[width=1\linewidth]{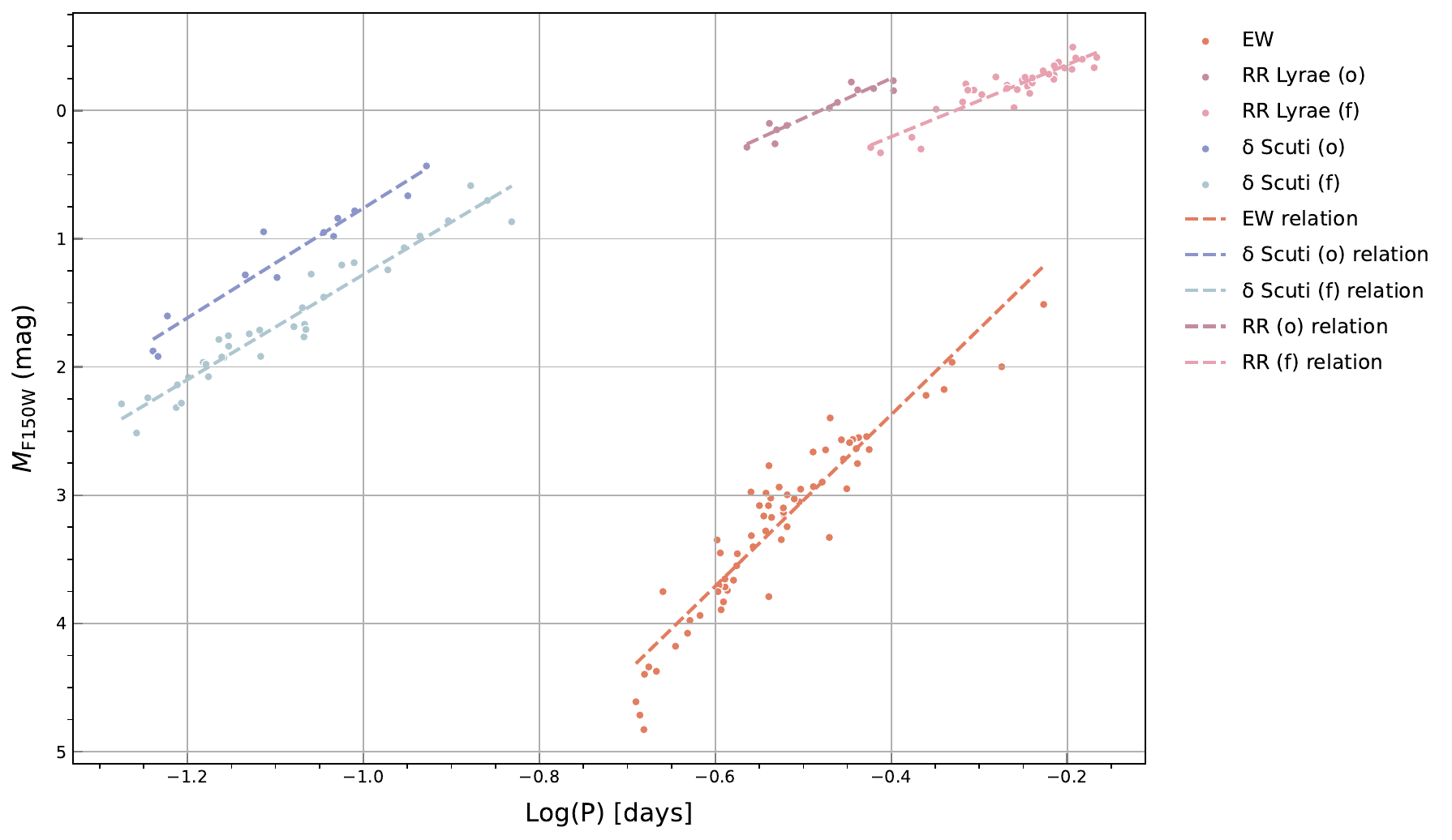}
    \caption{Distribution of EW-type eclipsing binaries, DSCT stars, and RR Lyrae stars in the period–luminosity diagram. Dashed lines indicate the best-fit PLR for each variable type. Different colors are used to distinguish between variable star types and, where applicable, between different pulsation modes (e.g., fundamental and first-overtone modes).}
    \label{fig:6}
\end{figure*}

\begin{figure*}[ht]
    \centering
    \includegraphics[width=1\linewidth]{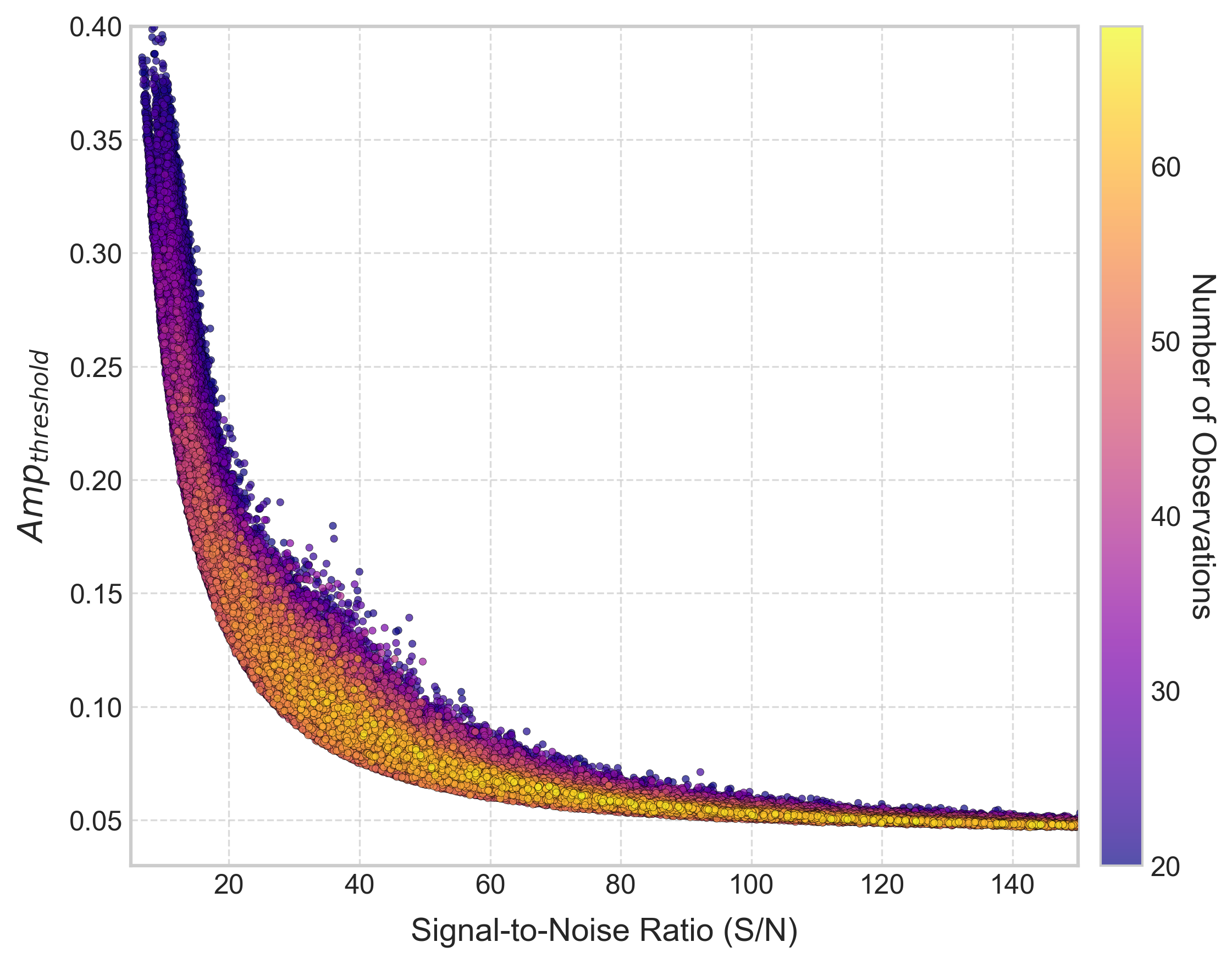}
    \caption{The figure presents the amplitude detection threshold for identifying variable stars as a function of SNRs in simulation. The results reveal a correlation with the amount of photometric data: the greater the number of measurements, the more sensitive the detection becomes to low-amplitude variability.}
    \label{fig:7}
\end{figure*}

\section{variable results}\label{sec:result}
This study performs cross-matching and integration of multi-source catalogs after completing astrometric corrections and photometric zero-point calibrations. Single-source samples with no fewer than 20 effective observations are selected for identifying periodic variable stars. We apply the Lomb-Scargle periodogram method to estimate the power spectral density \citep{1976Ap&SS..39..447L,1982ApJ...263..835S}. At the same time, the power spectrum of the observational window function is constructed to mitigate the impact of false periodicities introduced by sampling patterns. SNR thresholds of SNR $> 4$ are used to identify characteristic peaks in the window spectrum. Based on period uncertainty analysis, a frequency range of $\pm 0.0001$ c/d around each peak is defined as the window frequency interval. We applied the spectral power zeroing method to suppress the power values in the original power spectrum that fall within the window frequency interval, thereby effectively eliminating spurious periodic signals induced by the sampling pattern.

To determine the significance of periodic signals, this study adopts a false alarm probability (FAP) threshold of FAP $< 1 \times 10^{-5}$ to screen for statistically significant periodic signals. We select the inverse of the highest peak in the periodogram as the period estimate. After deriving the periods, a third-order Fourier series model is employed to perform phase-folded fitting of the LCs:
\[
f(t) = a_0 + \sum_{i=1}^3 a_i \cos\left(\frac{2\pi it}{P} + \phi_i\right)
\]
The goodness-of-fit is evaluated by the root mean square error (RMSE) and the coefficient of determination ($R^2$). Preliminary analysis reveals that significant pseudo-periodic signals persist even after window effect correction. Therefore, a dual-screening mechanism is established to improve the reliability: (1) Exclusion of systematic pseudo-signals based on period clustering effects: when frequency differences between different sources are  $ < 0.0001$ c/d, these are classified as systematic pseudo-signals, which appear as prominent vertical stripes in the period–power diagrams of the full sample; (2) Selection of high-confidence candidates, defined by amplitude (Amp) greater than 3 times the RMSE and $R^2 > 0.6$. We give priority to sources exhibiting consistent periods across multiple bands. Finally, after a limited visual inspection of the LC morphologies, 51 reliable variable stars are detected.

Although joint multi-band LC analysis can improve the reliability of candidate sources, it is limited by incomplete data coverage in some bands (only a few sources have multi-band repeated photometry) and the photometric precision constraints of certain bands, which may result in the loss of a large number of variable stars. Therefore, this study selects the F150W band, which offers the most complete photometric coverage, to carry out a single-band variable source search. Two criteria are set for single-band candidates: (1) High-amplitude sources (Amp $\geq 0.14$ mag, photometric scatter $\sigma \geq 0.04$ mag), ensuring that periodic signals are not overwhelmed by noise; (2) PLR-based screening. For candidate types with typical PLRs, such as DSCT stars, RR Lyrae stars, and EW-type eclipsing binaries, screening within the 5$\sigma$ range of their PLRs is performed, followed by visual inspection of LCs.

Regarding the classification of variable stars, the combination of CMDs and PLRs enables effective identification of DSCT stars and RR Lyrae stars, while sources with prominent eclipsing features are classified as eclipsing binaries. Classification between eclipsing binaries and rotational variables (ROT) is further refined using the phase parameter $\phi_{21} = \phi_2 - 2\phi_1$ \citep{2020ApJS..249...18C}: sources with $\phi_{21}$ values close to zero are classified as eclipsing binaries, while the rest are categorized as rotational variables. 

 In total, we identified 304 variable star candidates, including 77 eclipsing binaries, 12 RR Lyrae stars, 37 DSCT stars, and 178 rotational variables. Notably, the group labeled as ``rotational variables'' is presumably heterogeneous---this classification typically reflects cases where no specific variability type could be assigned with confidence. Therefore, it likely includes true rotational variables only as a subset. Cross-matching with the Gaia DR3 and OGLE-IV catalogs confirms the presence of 46 RR Lyrae and 16 DSCT stars in this region. Compared with these, our study newly identifies 4 RR Lyrae and 37 DSCT stars \citep{2016AcA....66..131S,2023AcA....73..105S,2023A&A...674A..18C}.

After the initial classification, we further refined the derived periods using iterative fitting. Given the sparse and irregular cadence of the photometric data, LS periodograms alone may lack sufficient precision. We therefore searched for the best-fit period within $\pm1\%$ of the initially derived period, using a fine step size of 0.00001 days. At each trial period, we performed phase folding and Fourier fitting, selecting the solution with the highest $R^2$ value. All folded light curves were visually inspected to avoid overfitting and to exclude spurious periods with artificially high $R^2$.

 Table~\ref{tab:1} presents 304 periodic variable stars identified through JWST near-infrared photometry in the LMC. The table includes the following parameters for each object: ID (serial number in the catalog), positions (J2000 RA and Dec), period, false alarm probability (FAP), mean magnitudes and associated uncertainties in six bands (F070W, F150W, F200W, F277W, F356W, F444W), number of detections, variability type, and LC parameters. Among these, only the F150W band provides the multi-epoch magnitude scatter $\sigma_{\mathrm{F150W}}$, while the other bands list only the mean magnitude and its estimated uncertainty.

The uncertainty of the mean magnitude, denoted as $m_{\rm err}$, is estimated using two methods depending on whether a reliable LC fit is available. For sources with a fitted light curve, we take the larger of the average single-epoch photometric error $\langle \Delta m \rangle$ and the RMSE of the model residuals, divided by the square root of the number of detections $N$:
\[
m_{\rm err} = \frac{\max(\langle \Delta m \rangle, \mathrm{RMSE})}{\sqrt{N}}.
\]
For sources without a valid LC fit, where RMSE is not available, we use the multi-epoch scatter in the F150W band, denoted as $\sigma_{\mathrm{F150W}}$, as a substitute for intrinsic variability. In this case, the uncertainty is estimated as:
\[
m_{\rm err} = \frac{\max(\langle \Delta m \rangle, \sigma_{\mathrm{F150W}})}{\sqrt{N}}.
\]

This substitution method is reasonable because, in the infrared, the decrease in pulsation amplitude with wavelength becomes very weak. Except for the F070W band, where the uncertainty may be underestimated by about 20\%. For eclipsing binaries and rotational variables, the variability amplitude is primarily caused by external factors and is largely independent of wavelength.

The table also includes LC parameters such as the total amplitude, Fourier amplitude ratio $R_{21}$, phase difference $\phi_{21}$, RMSE, and the coefficient of determination ($R^2$). Figure~\ref{fig:5} shows the distribution of these variable stars on the color--magnitude diagram (CMD), with non-variable stars displayed as a background density map for comparison. RR Lyrae stars are located along the horizontal branch, DSCT stars are distributed along the upper main sequence, and other types of variables appear more broadly scattered along the main sequence. Phased light curves in the F150W and F200W bands for multi-band sources are presented in the Appendix. For a few EW-type binaries, the amplitudes in F200W appear larger than those in F150W, which is likely attributable to sparse sampling and limited phase coverage, potentially leading to overfitting during Fourier fitting; photometric uncertainties may also contribute.

In addition, we provide the complete multi-epoch photometry catalog for all variable stars in Table~\ref{tab:2}, which includes the source ID, RA, Dec, filter, single-epoch magnitude, magnitude uncertainty, and observation time.

\section{Period-Luminosity Relations}\label{sec:plrs}
The PLR serves as a fundamental tool for studying periodic variable stars \citep{2020ApJS..249...18C}. It plays a crucial role both as a distance indicator and as a basis for variable star classification. In our sample of variable stars, including RR Lyrae stars, DSCT stars, and EW-type eclipsing binaries, well-defined PLRs are consistently observed.

To derive the PLR, we first corrected for interstellar extinction. We adopted the mean color excess $E(V-I) = 0.100 \pm 0.043$ reported by \citet{2021ApJS..252...23S}, and applied the extinction law for the LMC \citep{2019ApJ...877..116W, 2023ApJ...946...43W}. Accordingly, the average extinction values in the JWST near-infrared bands were calculated as follows:
A$_{F070W}$ = 0.163 mag, A$_{F150W}$ = 0.038 mag, A$_{F200W}$ = 0.021 mag, A$_{F277W}$ = 0.011 mag, A$_{F356W}$ = 0.007 mag, 
 A$_{F444W}$ = 0.004 mag. We obtained absolute magnitudes by subtracting the LMC distance modulus $\mu_0 = 18.477$ \citep{2019Natur.567..200P} from the extinction-corrected apparent magnitudes.

For eclipsing binaries, the Lomb–Scargle periodogram often detects half the orbital period \citep{2013AJ....146..101P}, Therefore, the derived periods were doubled when fitting the PLR for these systems. We prioritized the F150W and F200W bands due to their high photometric quality and completeness. To derive the PLR, we applied an iterative $3\sigma$ clipping method: sources with significant deviations from the initial PLR fit were excluded, followed by multiple rounds of fitting and outlier removal until convergence. Objects excluded in the PLR fit for either band were classified as EA-type binaries, while the remaining ones were classified as EW-type. In total, we identified 6 EA-type and 71 EW-type eclipsing binaries. It is important to note that our EA-type classification simply refers to sources that deviate significantly from the EW-type PLR, and in practice this group likely includes both detached and semi-detached eclipsing binaries. The EA:EW ratio in our sample is $\sim$1:12, which is substantially lower than the value of $\sim$3:1 reported by \citet{2025ApJS..276...57G} based on TESS data, $\sim$1:1.2 reported by \citet{2018MNRAS.477.3145J} based on the ASAS-SN all-sky survey and $\sim$1:7.4 reported by \citet{2020ApJS..249...18C} based on the ZTF survey. 
This discrepancy is primarily due to a selection effect introduced by our sparse and non-uniform photometric sampling: detached eclipsing binaries typically exhibit eclipses over a very small fraction of their orbital phase, making them less likely to be detected as periodic variables under such incomplete phase coverage. In contrast, EW-type eclipsing binaries exhibit nearly continuous light variation LC, allowing them to be identified more easily even with limited cadence observations. Considering the known difference in PLR between early- and late-type EW systems \citep{2018ApJ...859..140C}, we restricted the final PLR fit to late-type systems with $\log P_{\rm orb} < -0.25$~day. The fitting results shows a zero-point offset of approximately 0.1~mag relative to \citet{2018ApJ...859..140C}, while the slopes and intercepts are in excellent agreement with those reported by \citet{2025MNRAS.539..956L} across multiple bands (with zero-point differences less than 0.05 mag). That work utilized approximately 1000 EW-type eclipsing binaries with an average distance of 0.7 kpc to simultaneously determine the Gaia DR3 parallax zero-point and PLRs, achieving a PLR zero-point uncertainty of only 0.3\%. This consistency demonstrates that the PLRs of EW-type eclipsing binaries are highly robust and can be effectively used in future distance measurements to the LMC.

For pulsating variables, due to the relatively small number of sources discovered by JWST, we combined all DSCT and RR Lyrae stars (including those cross-matched with OGLE and Gaia) to derive a unified PLR fit. The period–luminosity diagram shows that, for both DSCT and RR Lyrae stars, the fundamental (F) and first-overtone (1O) pulsators occupy two distinctly separated sequences. Although these sequences are intrinsically distinct, they can be blended or smeared together in ground-based or lower-resolution space photometry due to source crowding in deep fields. Here, thanks to the extremely high spatial resolution of JWST, such blending effects are significantly mitigated, enabling a clear separation of the two pulsation modes in our PLR diagram. The PLR fitting results for all variable types, including EW-type eclipsing binaries, DSCT stars, and RR Lyrae stars in both F and 1O modes, are summarized in Table~\ref{tab:3}. Figure~\ref{fig:6} illustrates their distributions and fitted PLRs. The best-fit PLR scatter for RR Lyrae stars is 0.08 mag for the F mode and 0.04 mag for the 1O mode, while for DSCT stars, the scatter for both modes is 0.12 mag. For EW-type eclipsing binaries, the scatter is 0.22 mag.

Considering the higher precision of the PLR at longer wavelengths, we compared our F200W-band PLR with previous studies based on the 2MASS $K_s$-band. For DSCT stars, we found that the scatter in the fit was significantly reduced (from 0.17 mag to 0.12 mag). However, our zero-point is fainter by 0.15~mag compared to \citet{2025arXiv250415045L}, and by about 0.3~mag compared to \citet{2025arXiv250320557J} and \citet{2020MNRAS.493.4186J}. This discrepancy may arise from the following factors: (1) the effective wavelength of F200W ($\lambda_{\rm eff} = 1.97\ \mu$m) is shorter than that of the $K_s$ ($\lambda_{\rm eff} = 2.16\ \mu$m), and it is known that for DSCT stars the zero-point decreases with increasing wavelength from the optical to the near-infrared; (2) JWST offers excellent spatial resolution in the near-infrared ($\sim$0.05$^{\prime\prime}$ in the near-infrared), significantly mitigating crowding effects; (3) issues in Gaia parallax zero-point determinations — \citet{2025arXiv250320557J} (zp = 56.2~$\mu$as) and \citet{2025arXiv250415045L} (zp = 35~$\mu$as) reported relatively large zero-point offsets. Compared to Cepheids, DSCT stars are approximately 3–7 magnitudes fainter, and compared to EW-type eclipsing binaries, DSCT stars are about an order of magnitude fewer in number within the same volume. Therefore, DSCT stars are not well suited for simultaneously calibrating the Gaia parallax zero point and deriving reliable PLRs due to the limited sample. Instead, constraining the DSCT PLR using the distance modulus of the LMC yields more accurate results. This comparison indicates that the DSCT PLR still has room for improvement.

Similarly, our PLR for RR Lyrae stars shows a zero-point offset of only $\sim$0.05~mag relative to the $K$-band PZL relation of \citet{2023ApJ...951..114Z}, when metallicity effect is not considered. This level of discrepancy is within expectations. Since RR Lyrae stars are intrinsically brighter than DSCT stars, they are less affected by crowding.

If we construct the PLRs of the same RR Lyrae and DSCT star samples using OGLE or Gaia magnitudes, the resulting dispersions are more than twice as large as those obtained in this study. This strongly indicates that the OGLE and Gaia magnitudes for these RR Lyrae and DSCT stars are significantly affected by crowding effects. Regarding the crowding effects, we conducted a check in our field. By cross-matching OGLE DSCT sources with our JWST F150W catalog, we found that each OGLE source typically has more than five contaminating sources within a 0.4$^{\prime\prime}$ radius, resulting in an average systematic brightening of approximately $\sim$0.38 mag. Similarly, we tested RR Lyrae stars, and found that contamination leads to an average brightening of about $\sim$0.12 mag, consistent with predictions based on the differences in absolute magnitudes. The region we studied is one of the denser fields in the LMC; similar crowding is expected in other LMC regions and in Galactic disk samples, although the impact would be smaller. 

\section{discussion}\label{sec:discuss}
Based on the CMD distribution shown in Figure~\ref{fig:5}, the F150W absolute magnitudes of stars in our sample span a range of $-1.5$ to 7~mag. Given the diverse stellar populations in the LMC, the sample is theoretically expected to include pulsating variables such as slowly pulsating B Stars (SPB), DSCT stars, $\gamma$ Doradus (GDOR) stars, and RR Lyrae stars, as well as rotational variables (e.g., Be stars and magnetic variables) and numerous eclipsing binaries. However, the sparse and inhomogeneous sampling frequency of JWST observations restricts detectable variability to short-period variables with relatively high amplitudes.

Considering the wavelength-dependent amplitude attenuation of pulsating variables (near-infrared amplitudes being significantly smaller than their optical counterparts), only a limited number of pulsators meet the detection criteria. The minimum amplitude observed in our sample is $\mathrm{Amp}_{F150W} \approx 0.05$~mag, whereas literature studies indicate typical amplitudes for SPBs below 0.02~mag \citep{2023AcA....73..105S}, and non-radial pulsation amplitudes for DSCT stars ranging from a few to tens of millimagnitudes \citep{2025ApJ...978...53W}. Consequently, only high-amplitude RR Lyrae stars and high-amplitude DSCT stars are likely detectable. This expectation is borne out by our results, which show that JWST’s variable-star detection capability is effectively confined to rotational variables, eclipsing binaries whose amplitudes are minimally affected by wavelength attenuation, and a small subset of high-amplitude pulsators.

Under ideal conditions (non-variable assumption with infinite photometric measurements), the photometric scatter of a star should equal its photometric error. However, practical limitations—including finite sampling and systematic uncertainties in photometric errors—significantly complicate variability detection. For DSCT stars in our sample, the SNR exceeds 100 (theoretical photometric error $<0.01$~mag), yet the observed minimum amplitude remains 0.05 mag, implying the presence of unaccounted errors.

Photometric error analysis reveals detector-dependent zero-point offsets of 0.01--0.2~mag in JWST~\citep{2022RNAAS...6..191B}, with residual systematic deviations $<0.04$~mag even after standard star calibration. Additional dispersion introduced by PSF typically contributes 0.01 mag and can reach up to 0.07 mag in extreme cases \citealt{2024ApJS..271...47W}). Furthermore, photometric degradation due to spacecraft jitter exacerbates the noise floor. To quantify the impact of these noise sources, we introduce a systematic error compensation of 0.02 mag to the photometric uncertainties and establish a Monte Carlo simulation framework. Assuming that the observed scatter of non-variable stars is entirely dominated by photometric errors, we perform 10,000 resamplings for each star, using the individual photometric error $\Delta m_i$ of each data point as the sampling parameter. Each resampling generates a new set of photometric measurements and its corresponding scatter. Thus, the 10,000 resamplings result in a distribution of scatter for each source. The median of this distribution converges to the photometric error, and as the number of photometry increases, the scatter distribution narrows, implying that the error-induced scatter will converge to the photometric error. However, when the number of observations is limited, the total scatter $\Delta m_{total}$ can substantially exceed the nominal per-point error. 

For variable stars, observed scatter $\sigma$ is expected to exceed $\Delta m_{total}$ due to intrinsic variability. Typically, a total amplitude greater than twice the scatter $\sigma$ of non-variable stars is more likely to indicate genuine variability. In our analysis, we find that the amplitudes of all identified variable stars in our sample exceed twice the $(\mathrm{median} + 1\sigma)$ quantile of the simulated scatter distribution. Based on this, we define a star as significantly variable if its scatter $\sigma$ exceeds twice the 84th percentile of the simulated distribution. We further utilize this simulation framework to assess JWST's sensitivity to small-amplitude variability. As shown in the Figure~\ref{fig:7}, the simulations results indicate that even under high SNR conditions, a variability amplitude of approximately 0.05 mag is required to rise above the noise floor and be detected. This threshold matches the smallest amplitudes observed in our data and confirms the presence of systematic uncertainties at the level of $\sim$0.02 mag, which are not accounted for by nominal photometric errors. Increasing the observational cadence substantially reduces the minimum detectable amplitude, particularly for low-SNR sources. For instance, at $\mathrm{SNR} = 20$, the amplitude threshold decreases from approximately 0.2 mag with 20 exposures to about 0.12 mag with 60 exposures.. 

These findings explain the composition of detected variable stars: (1) Eclipsing binaries and rotational variables dominate the sample, as their amplitudes in the near-infrared are comparable to those in the optical; (2) The detectability of DSCT stars is limited due to amplitude attenuation in the near-infrared and the intrinsically low fraction of high-amplitude DSCT stars ($<5\%$; \citealt{2020MNRAS.499.3976P}); (3) Brighter RR Lyrae stars are more readily detected due to their larger amplitudes and higher SNRs. Our analysis defines the sensitivity limits of JWST in variable star detection and provides critical guidance for optimizing future observing strategies, including increasing sampling rates and extending monitoring durations.

\section{conclusion}\label{conclude}
This study evaluates the feasibility of periodic variable star detection with JWST by analyzing multi-epoch photometric data of the LMC using DOLPHOT-based PSF photometry, combined with CMD diagnostics. Prior to variable star identification, we applied astrometric corrections and photometric zero-point alignments to ensure accurate cross-epoch source matching. Through LC analysis, we identified a total of 304 variable stars--most of which are newly discovered--including 71 EW-type eclipsing binaries, 6 EA-type eclipsing binaries, 178 rotational variables, 37 DSCT stars, and 12 RR Lyrae stars. We provide a catalog that includes periods, amplitudes, and mean magnitudes in six JWST/NIRCam filters: F070W, F150W, F200W, F277W, F356W, and F444W.

To investigate the PLRs, we cross-matched our photometric sample with the OGLE and Gaia catalogs to increase the number of RR Lyrae and DSCT stars. In contrast, the PLR analysis of EW-type eclipsing binaries was conducted exclusively using our own newly identified sources, independent of external catalogs. The PLR results show that the best-fit $1\sigma$ scatter for RR Lyrae stars is 0.08 mag for the F mode and 0.04 mag for the 1O mode; for DSCT stars, the scatter for both modes is 0.12 mag; and for EW-type eclipsing binaries, it is 0.22 mag. Compared to previous studies, the PLR zero point for DSCT stars appears systematically fainter, which can be attributed to the shorter effective wavelength of JWST's F200W filter relative to the 2MASS $K_{s}$ band, as well as JWST's superior spatial resolution, which effectively mitigates crowding effects. Compared to earlier PLR studies, our work benefits from two key advantages: first, JWST’s high spatial resolution enables more accurate photometry in crowded stellar fields, significantly reducing the effects of crowding; second, the LMC provides a well-determined distance that is independent of parallax, offering a robust distance anchor for PLR calibration. Through crowding effect analysis, we find that in this region, the magnitudes of RR Lyrae and DSCT stars from OGLE and Gaia are affected by approximately 0.38 and 0.12 mag, respectively. Therefore, great caution is required when using OGLE and Gaia data to study RR Lyrae and DSCT stars in the LMC.

Due to the fact that most JWST observing programs do not perform long-term, continuous, and uniform sampling in the same sky region, only variable stars with relatively large amplitudes are typically detectable. Systematic effects—such as telescope jitter, time-varying PSF characteristics \citep{2024ApJS..271...47W}, and residual detector-to-detector zero-point offsets \citep{2022RNAAS...6..191B}—further degrade photometric precision beyond the formal error estimates provided by DOLPHOT. To account for these effects, we introduced an additional 0.02 mag systematic uncertainty to the photometric errors and conducted Monte Carlo simulations, constrained by the properties of the variable stars we identified. The simulations demonstrate that the detectability of low-amplitude variables is highly sensitive to both the SNRs and the number of observational epochs. For sources with SNR$\sim20$, the minimum detectable amplitude decreases from 0.2 mag with 20 exposures to 0.12 mag with 60 exposures, emphasizing the importance of sampling cadence in variable star discovery.

JWST remains highly effective in detecting short-period, large-amplitude variables, particularly eclipsing binaries, rotational variables, and high-amplitude pulsators. Moreover, JWST’s high-precision photometry can significantly improve the measurements of known variables, especially for calibrating PLRs in the near-infrared. Its exceptional spatial resolution greatly mitigates crowding effects in dense stellar fields, enabling more accurate photometric measurements. As JWST continues to accumulate additional epochs, its unparalleled sensitivity and spatial resolution in the near-infrared will play an increasingly important role in extragalactic variable star studies.

\section*{Acknowledgements}
We thank the anonymous referee for the valuable comments. This work was supported by the National Natural Science Foundation of China (NSFC) through grants 12173047, 12322306, 12373028, 12233009 and 12133002. We also thanked the support from the National Key Research and development Program of China, grants 2022YFF0503404. X. Chen and S. Wang acknowledge support from the Youth Innovation Promotion Association of the Chinese Academy of Sciences (CAS, No. 2022055 and 2023065). This work is based on observations made with the NASA/ESA/CSA James Webb Space Telescope. The data were obtained from the Mikulski Archive for Space Telescopes (MAST) at the Space Telescope Science Institute.

\software{Astropy \citep{astropy:2013, astropy:2018, astropy:2022}}, Matplotlib \citep{2007CSE.....9...90H}, TOPCAT \citep{2005ASPC..347...29T}, DOLPHOT \citep{2016ascl.soft08013D}

\bibliography{sample631}
\bibliographystyle{aasjournal}

\section{Appendix}\label{appendix}

\begin{figure*}[ht]
    \centering
    \includegraphics[width=1\linewidth]{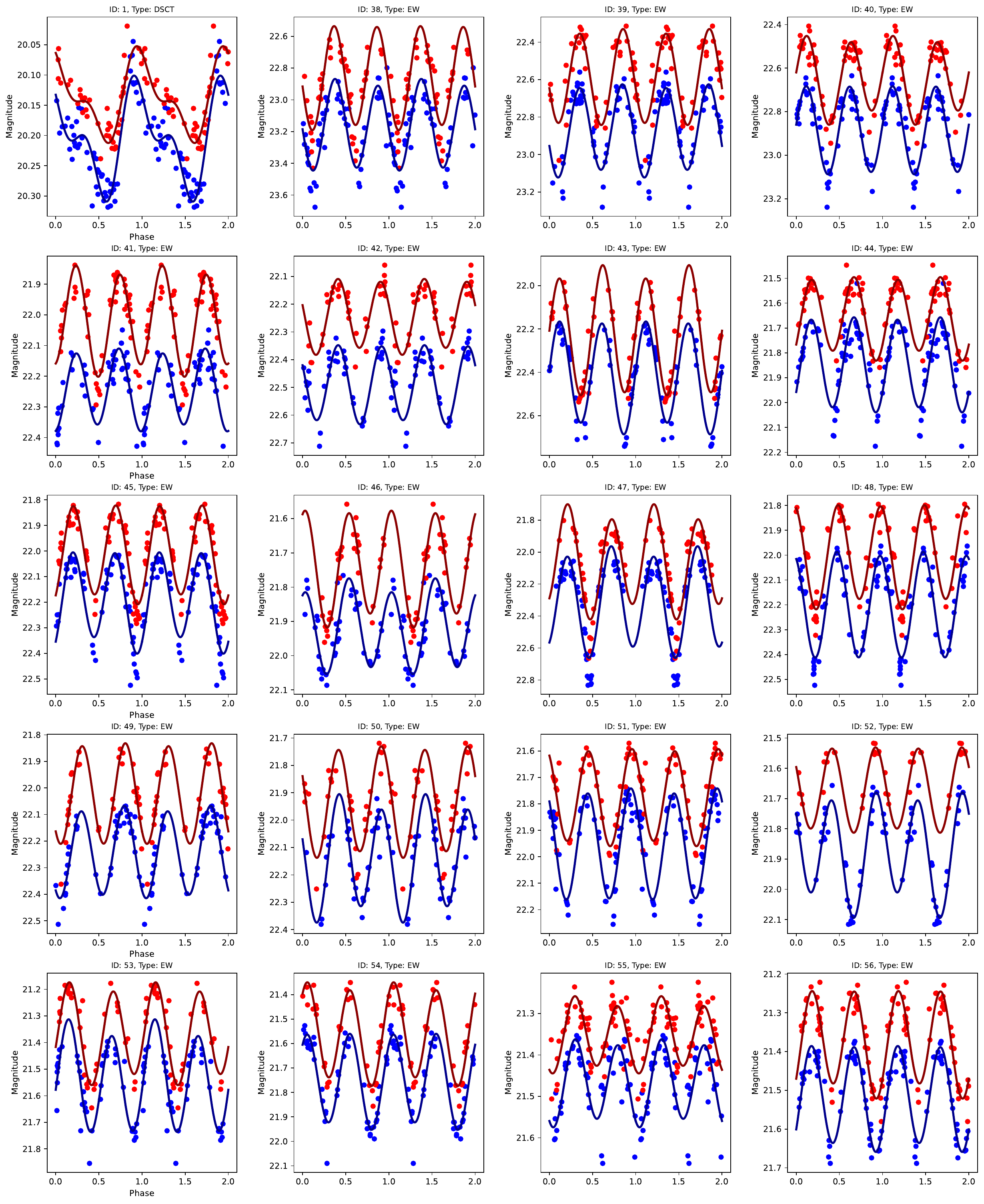}
    \caption{ Phase-folded LCs of variable stars exhibiting consistent periods across multiple bands. Blue and red curves correspond to observations in the F150W and F200W filters, respectively. For eclipsing binaries, the curves are folded using twice the derived photometric period to represent the full orbital cycle.}
    \label{A1}
\end{figure*}

\begin{figure*}[ht]
    \centering
    \includegraphics[width=1\linewidth]{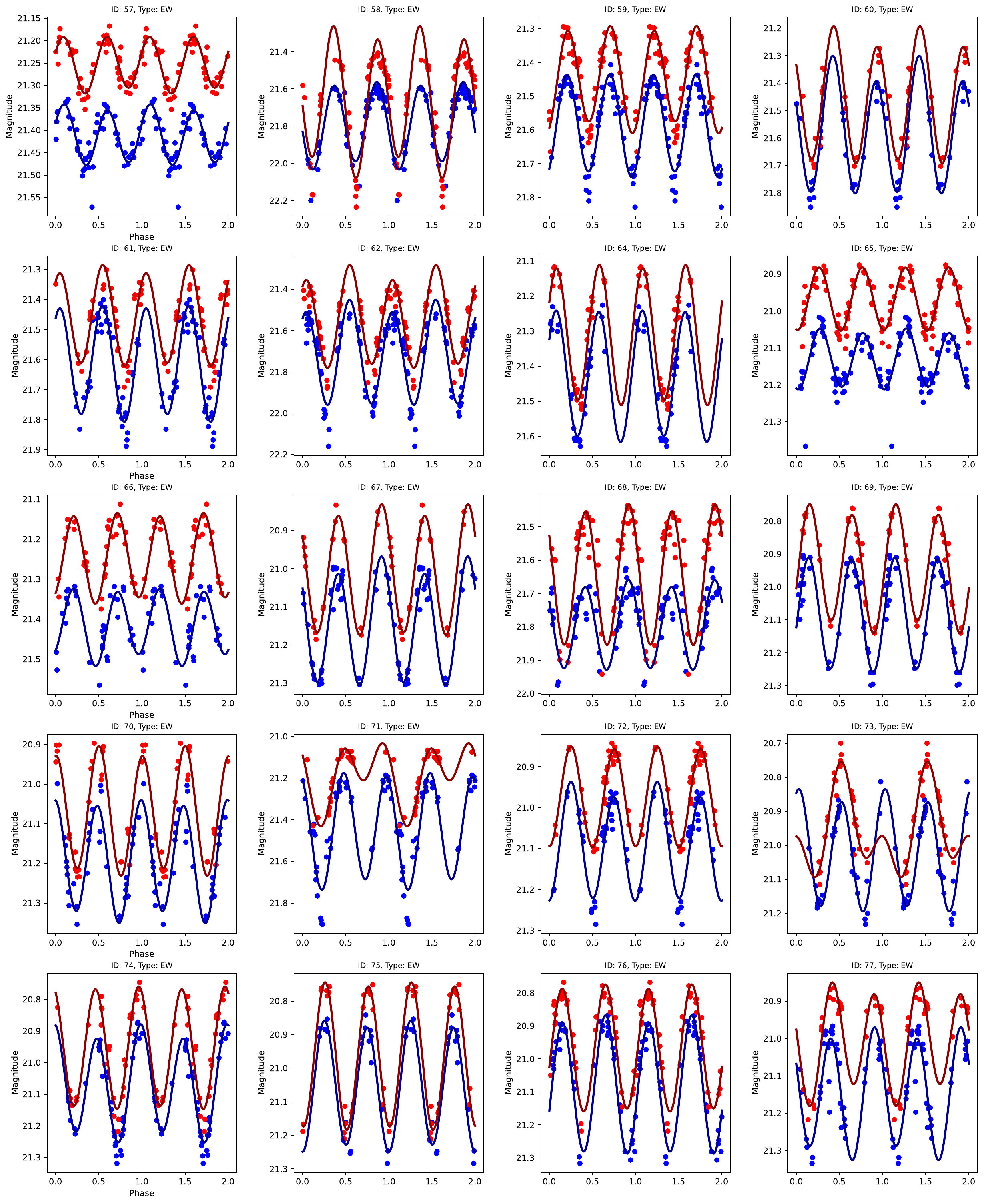}
    \caption{Similar to Figure~\ref{A1}.}
    \label{A2}
\end{figure*}

\begin{figure*}[ht]
    \centering
    \includegraphics[width=1\linewidth]{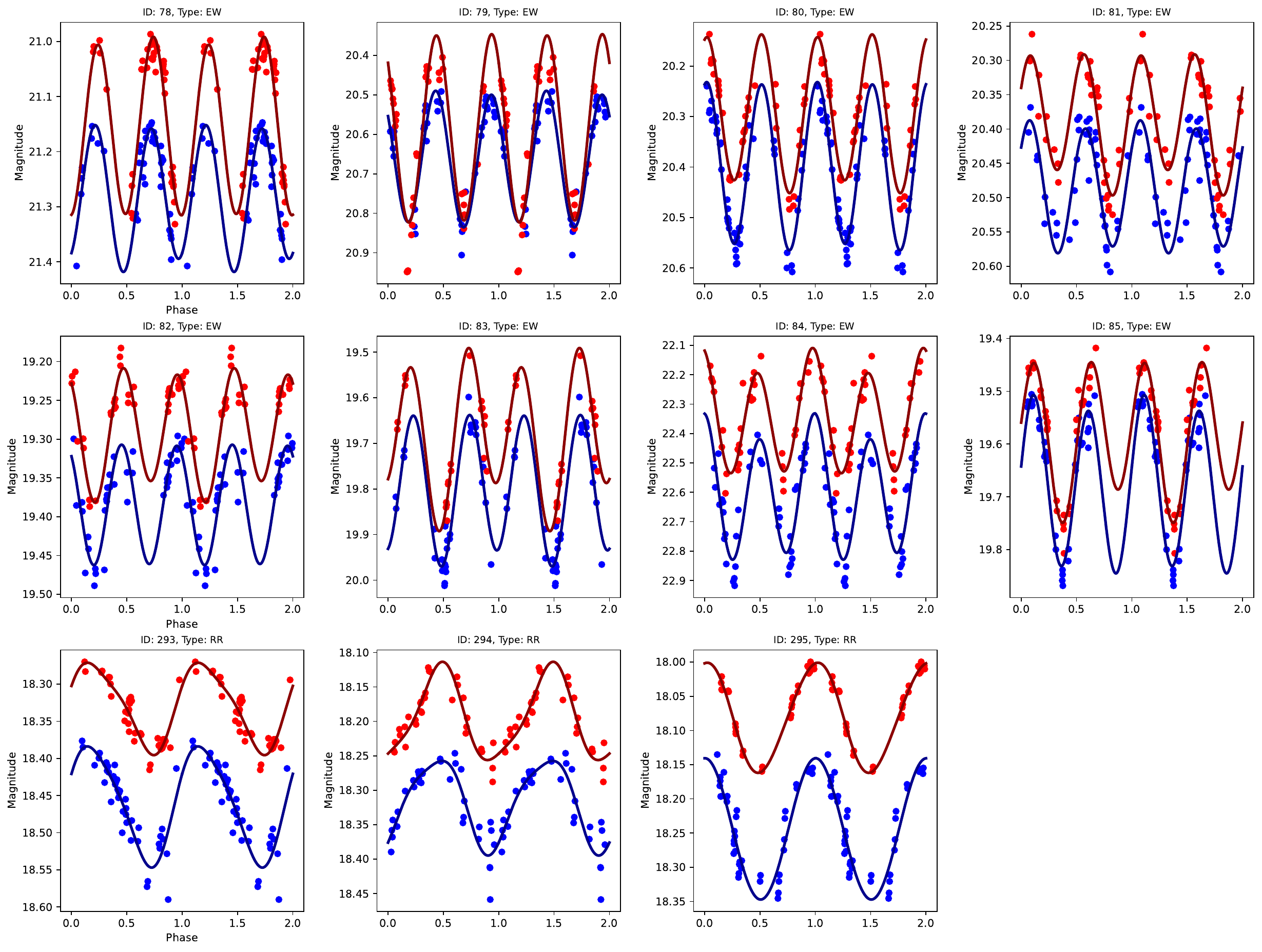}
    \caption{Similar to Figure~\ref{A1}.}
    \label{A3}
\end{figure*}

\begin{figure*}[ht]
    \centering
    \includegraphics[width=1\linewidth]{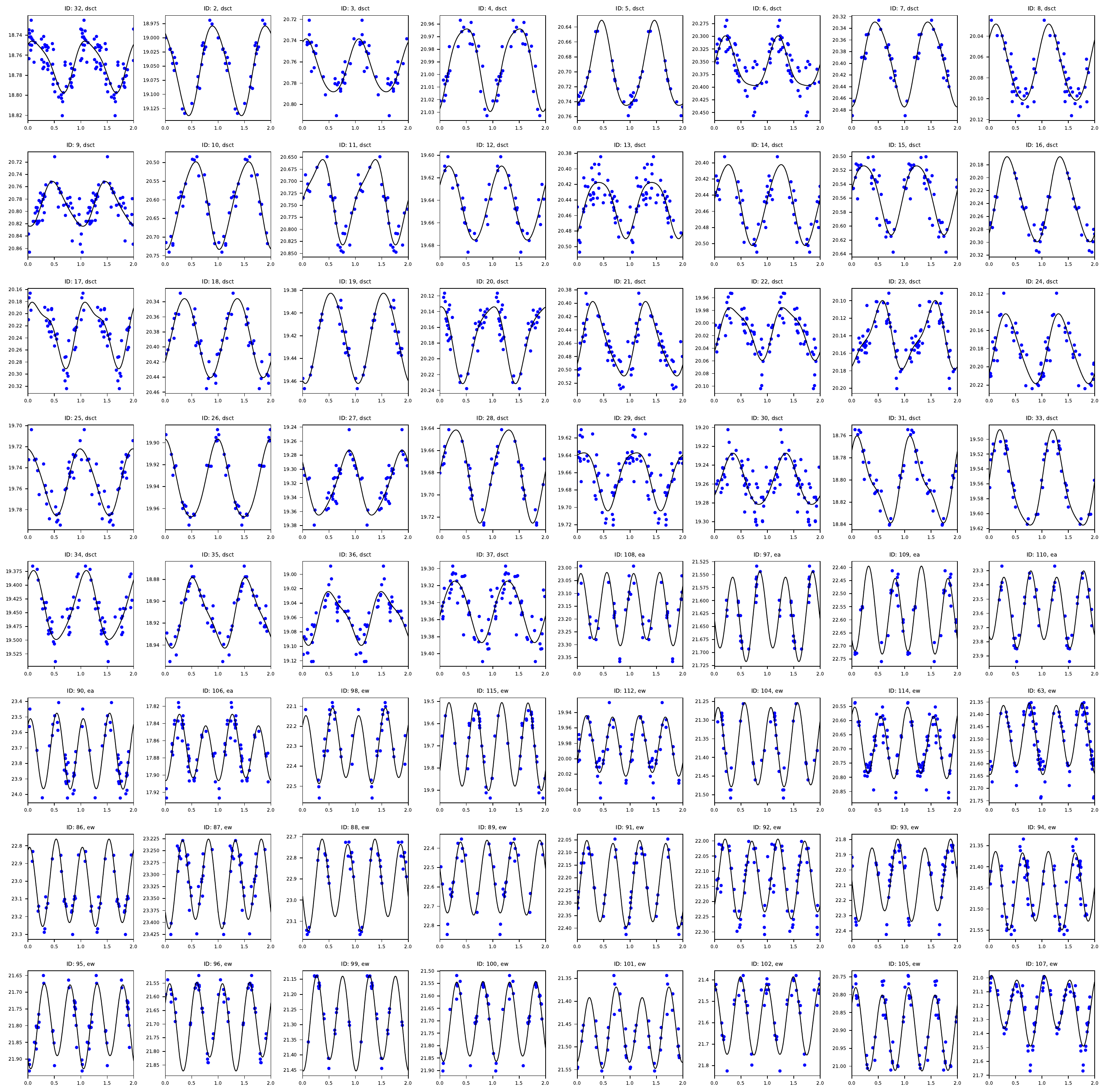}
    \caption{Phase-folded light curves of variable stars observed in the F150W filter. }
    \label{A4}
\end{figure*}

\begin{figure*}[ht]
    \centering
    \includegraphics[width=1\linewidth]{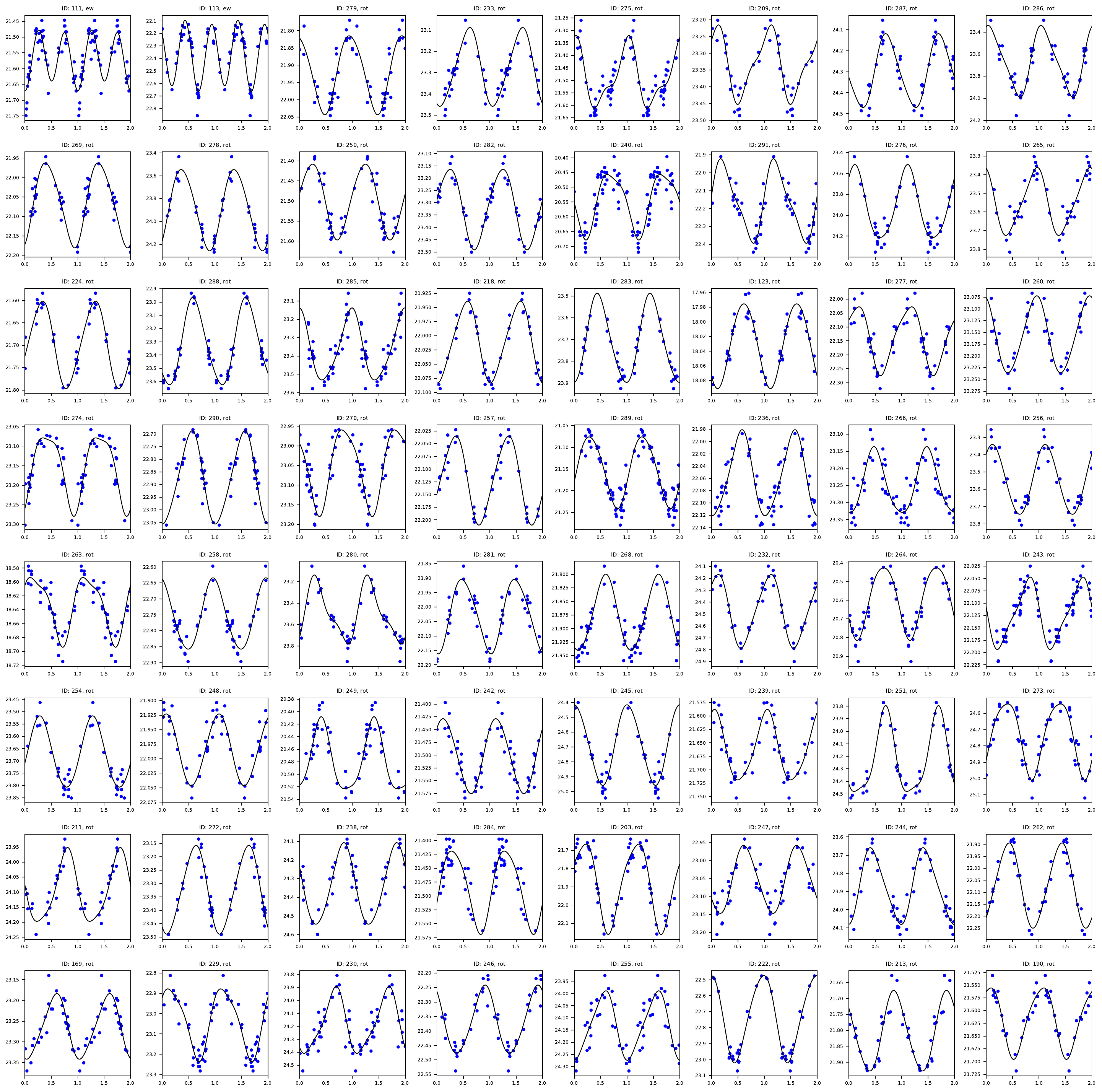}
    \caption{Similar to Figure~\ref{A4}. }
    \label{A5}
\end{figure*}

\begin{figure*}[ht]
    \centering
    \includegraphics[width=1\linewidth]{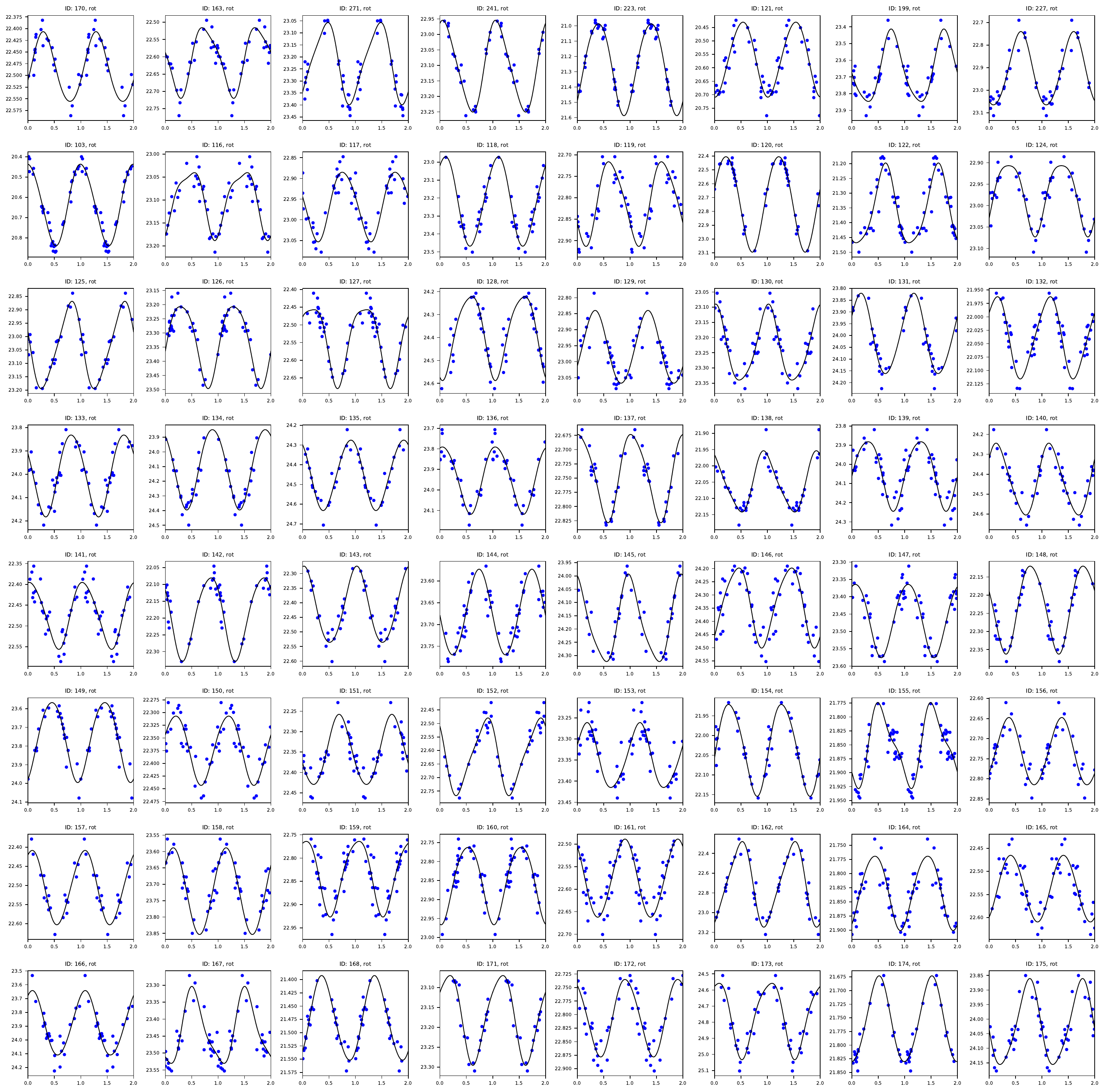}
    \caption{Similar to Figure~\ref{A4}. }
    \label{A6}
\end{figure*}

\begin{figure*}[ht]
    \centering
    \includegraphics[width=1\linewidth]{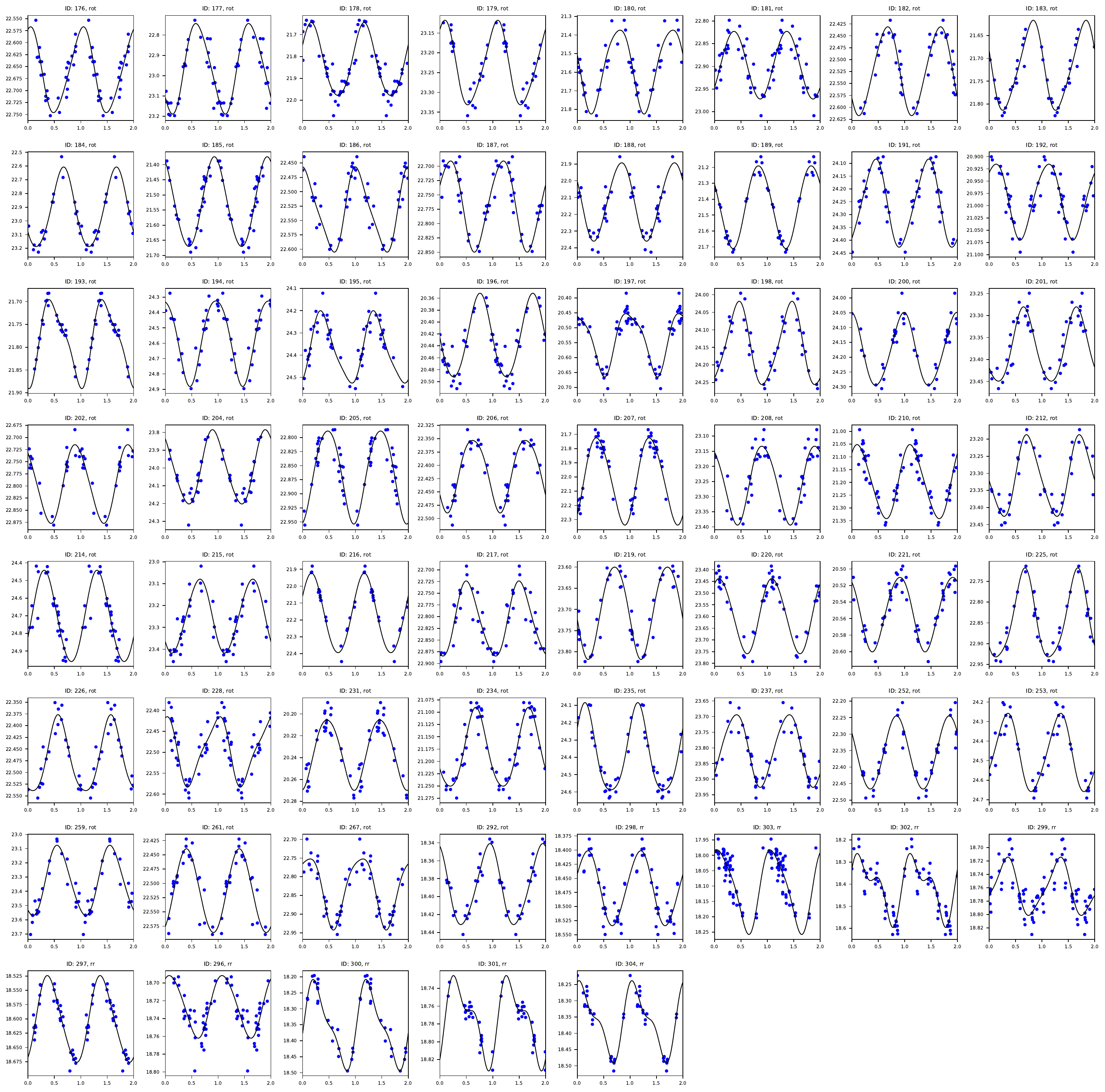}
    \caption{Similar to Figure~\ref{A4}. }
    \label{A7}
\end{figure*}

\end{document}